\documentclass[prb,twocolumn,longbibliography,superscriptaddress, floatfix]{revtex4-2}
\usepackage{graphicx}
\usepackage{textgreek}

\usepackage{amsmath}
\usepackage{amssymb}
\usepackage{hyperref}
\usepackage[caption=false]{subfig}
\usepackage[capitalize]{cleveref}
\usepackage{braket}
\usepackage{mathtools}
\usepackage{interval}
\usepackage{bm}
\intervalconfig{soft open fences}

\AddToHook{cmd/appendix/before}{
    \crefalias{section}{appendix}
    \crefalias{subsection}{appendix}
}

\AtBeginDocument{
  \renewcommand{\Re}{\operatorname{Re}}
  \renewcommand{\Im}{\operatorname{Im}}
}

\DeclareMathOperator{\tr}{tr}

\DeclarePairedDelimiter\abs{\lvert}{\rvert}

\crefname{section}{Sec.}{Sections}

\begin{document}
\title{Superconductivity via paramagnon and magnon exchange in a 2D
    near-ferromagnetic full metal and ferromagnetic
    half-metal}

\author{Zachary M. Raines}
\author{Andrey V. Chubukov}
\affiliation{School of Physics and Astronomy and William I.
    Fine Theoretical Physics Institute, University of Minnesota, Minneapolis, MN 55455, USA}
\begin{abstract}
We study superconductivity in paramagnetic and ferromagnetically-ordered phases in a two-dimensional electron system with parabolic fermionic dispersion and short-range repulsive interaction.
In the paramagnetic phase, we find that a weak momentum dependence of a paramagnon propagator parametrically reduces the onset temperature for the pairing compared to that in phenomenological theories which assume a strong dispersion of a paramagnon
and also changes the topology of the gap function.
In the ferromagnetic phase, we show that the order instantly polarizes low-energy fermionic excitations.  We derive the fully renormalized pairing interaction between low-energy fermions, mediated by two transverse Goldstone modes and show that it is attractive in a spatially-odd channel.
The pairing temperature in the ferromagnetic phase is found to be a fraction of the Fermi energy, significantly larger than in the paramagnetic phase near the transition.
Our results are relevant for understanding superconductivity in proximity to itinerant ferromagnetism in multi-valley graphene systems, particularly the ones with full valley and spin polarization.
\end{abstract}

\maketitle

\section{Introduction}
\label{sec:intro}

Superconductivity mediated by ferromagnetic fluctuations has been discussed as far back as 1966 when Berk and Schrieffer~\cite{Berk1966} suggested that, while magnetic fluctuations were detrimental to s-wave superconductivity,
they do support an unconventional triplet pairing.
This idea was further explored in the context of $^3$He by
Layzer and Fay~\cite{Layzer1971} (see Ref.~\onlinecite{Leggett1975} for a review).
Subsequent studies focused on the interplay between superconductivity and non-Fermi liquid behavior, also caused by ferromagnetic spin fluctuations~\cite{Rosch2007,Chubukov2020a}, pair-breaking effects by thermal spin fluctuations~\cite{Monthoux1999,Wang2001,Roussev2001,Chubukov2003}, which for spin-triplet pairing act as magnetic impurities~\cite{AG1961,Littlewood1992}, and the non-analytic terms in the expansion of the Free energy over magnetization~\cite{Belitz2005,Chubukov2004,*Maslov2009,Brando2016}.
An outcome of these studies is that triplet superconductivity does develop and $T_c$ generally increases as the system approaches an instability towards ferromagnetism, but at large magnetic correlation length the transition may become first order.

Less is known about superconductivity coexisting with a magnetic order, although this issue
was also discussed
by Fay and Appel back in 1980~\cite{Fay1980}.
It is tempting to assume that the superconducting region forms a dome above the onset point of ferromagnetism at $T=0$ (a ferromagnetic quantum-critical point (QCP)), i.e., that superconductivity decreases roughly symmetrically upon deviations from a QCP into the paramagnetic
or magnetically ordered phases.
A near-symmetric dome-like shape of the superconducting region has been experimentally detected~\cite{Mathur1998} in itinerant antiferromagnets, like CePd$_2$Si$_2$.
However, in three-dimensional (3D) itinerant ferromagnets, such as UGe$_2$~\cite{Saxena2000}, URhGe~\cite{Aoki2001},
UCoGe \cite{Huy2007},
a superconducting region is located largely within a magnetically ordered phase, where
superconductivity co-exists with ferromagnetism.
A similar behavior has been detected in recent studies of
two-dimensional (2D) twisted bilayer graphene (TBG) (see e.g., Ref.~\onlinecite{Kozii2022} and references therein)
and non-twisted graphene-based materials, like Bernal-stacked bilayer graphene (BBG), rhombohedral  tri-layer and
penta-layer
graphene  (RTG and R5G, respectively) in the presence of a displacement field and, in some cases, Ising spin orbit coupling, induced by placing WSe$_2$ near graphene sheets~\cite{Seiler2022,Zhou2022a,Zhou2021,DeLaBarrera2022,Seiler2023,Arp2023,Holleis2023,Zhang2023a,Blinov2023,*Blinov2023a,
    Patterson2024,Han2025}.
In BBG and RTG,
superconductivity has been detected within a half-metal state~\cite{Holleis2023,Zhang2023a,Patterson2024},
which was experimentally identified as a ferromagnetic state~\cite{Patterson2024}, and in  R5G superconductivity has been detected~\cite{Han2025} inside a quarter-metal state, which likely has both ferromagnetic and valley order.

On the theory side,
several groups~\cite{Layzer1971,Levin1978,Blagoev1999,Wang2001,Roussev2001} computed $T_c$ in the paramagnetic phase in 3D due to exchange of ferromagnetic spin fluctuations (paramagnons).
Fay and Appel~\cite{Fay1980} computed $T_c$ in 3D both
in the paramagnetic phase and in the ferromagnetic phase, due to the exchange of spin-conserving longitudinal fluctuations, and found a near-symmetric superconducting dome around a ferromagnetic QCP.\@
Kirkpatrick et al.~\cite{Kirkpatrick2001} extended these calculations to include the coupling between longitudinal and transverse (Goldstone) modes and argued that this gives rise to the strong enhancement of $T_c$ in the ferromagnetic phase, compared to
the paramagnetic phase.
In 2D, several groups computed superconducting $T_c$ in the paramagnetic phase~(see e.g., Refs.~\onlinecite{Monthoux1999,Chubukov2003}).
Superconductivity in the ferromagnetically-ordered state has been recently analyzed in
magic-angle TBG~\cite{Kozii2022}, BBG/RTG with Ising-like spin-orbit coupling~\cite{Dong2024} and
in R5G~(Refs.~[\onlinecite{Chou2025,*Geier2024,*Yang2024,*Qin2024,*Jahin2025,*Gil2025,*Dong2025, *Gaggioli2025,*Yang2024, *Parramartinez2025,*Kim2025,*Christos2025}]).
Several groups also analyzed proximity-induced superconductivity in a metal placed next to a magnetically ordered insulator (see [\onlinecite{Maeland2023}] and references therein).

In this communication we present a comparable analysis of superconductivity on both sides of a ferromagnetic QCP in a 2D metal, using a specific microscopic model of fermions occupying a single valley, with a parabolic dispersion and Hubbard-like interaction at small momentum transfer.
We derive the spin-mediated pairing interaction both in the normal and
ferromagnetic state within the ladder approximation, by summing up infinite series of ladder diagrams containing particle-hole polarizations.
Within this approximation, ferromagnetic order appears as a result of a Stoner instability at $U \nu=1$, where $\nu =m/(2\pi)$ is the fermionic density of states per spin.
In 2D, this transition is
strongly first order, going directly to a state with full spin polarization~\cite{Raines2024a,*Raines2024b,*Raines2024c}, yet the static susceptibility diverges as the transition is approached from the paramagnetic side.
Our goal is to understand how this affects superconductivity near the Stoner transition.

Our key results can be briefly summarized as follows.
In the paramagnetic phase, we obtain the commonly used
Ornstein-Zernike + Landau damping form of the dynamical spin susceptibility, however with a small prefactor for the static gradient $q^2$ term.  We find that this smallness \textit{reduces}\/ the strength of the attractive pairing interaction in the spin-triplet channel, reduces the magnitude of $T_c$ and  changes its dependence on the ratio $U/E_F$, where $E_F$ is the Fermi energy.
It also gives rise to a topologically non-trivial gap function with the sign change on the Matsubara axis.
In the ferromagnetically-ordered state, we find an attraction from the process involving two gapless magnons.
We show that this process gives rise to larger $T_c$ than in the paramagnetic phase
even though the corresponding vertex function is reduced in agreement with the Adler principle for an interaction between fermions and a Goldstone boson.
We argue that superconductivity around a ferromagnetic QCP in 2D is largely confined to the region where it co-exists with full spin polarization.

As we will be primarily interested in the comparison of pairing scales in the paramagnetic and ferromagnetic phases,
we
restrict our discussion
to zero temperature where the Mermin-Wagner theorem does not hold, and long range order is possible.
While finite temperature effects are certain to be quantitatively important, they should not, to first order, modify the qualitative comparison of the pairing scales.

The outline of the paper is as follows.
In \cref{sec:model} we introduce the model and review the nature of the two-dimensional Stoner transition.
In \cref{sec:paramagnon-pairing} we consider pairing mediated by paramagnons in the paramagnetic phase.
In \cref{sec:magnon-pairing} we consider pairing mediated by Goldstone modes in the ferromagnetic phase.
In \cref{sec:conclusion} we summarize our results and discuss their application to 2D materials,
including R5G.\@

\section{Model and ferromagnetic order}\label{sec:model}

We
consider a 2D model of spinful fermions in two-dimensions.
Having in mind applications to BBG, RTG and R5G, we assume that fermionic density is low and approximate fermionic dispersion by a parabola $k^2/(2m)$.
We define the Fermi energy as $E_F = k_F v_F/2$ and assume that it is smaller than the bandwidth $W$.
We assume a contact repulsive interaction $U$ between electrons up to a certain momentum transfer $q_c \sim (2m W)^{1/2} > k_F$,
and set $U=0$ at larger momentum transfers.
The model is described by the Matsubara action
\begin{align}
    S  = {} & \sum_{k}\bar{\psi}_{k\sigma}\left(-i\omega_{n} + \frac{k^{2}}{2m} -\mu \right)\psi_{k\sigma} \nonumber \\
    + {}    & \frac{U}{2}\sum_{kk'q}\bar{\psi}_{k+q\sigma}\psi_{k\sigma}\bar{\psi}_{k'\sigma'}\psi_{k'+q\sigma'},
    \label{eq:S-model}
\end{align}
where $k = (\mathbf{k}, i\omega_n)$ combines fermionic momentum and Matsubara frequency, and $\sigma$ denotes spin.
We will measure the effects of $U$ in terms of the dimensionless coupling
\begin{equation}
    c =  U \nu
\end{equation}
where $\nu$ is the density of states ($=m/(2\pi)$ for a parabolic dispersion $\epsilon_k = k^2/(2m)$).

We will analyze the model of~\cref{eq:S-model} within the ladder approximation, i.e., will obtain the propagator for ferromagnetic spin fluctuations and the effective 4-fermion interaction, mediated by these fluctuations, by summing series of ladder and bubble diagrams.
Within this approximation, at $T=0$,
the system exhibits a Stoner quantum phase transition between a paramagnetic metal (PM) and a ferromagnetic metal (FM) at $c=1$.
The self-consistent equation for the ferromagnetic order parameter $\Delta$ is
\begin{equation}
    \Delta = \frac{U}{2} T\sum_{k} \tr[\hat{G}_{k}\hat{\sigma}^{z}],
\end{equation}
where $\hat{G}$ is a two-component diagonal matrix Green's function for spin-up and spin-down fermions.
This
equation must be supplemented by the
condition on the total density
\begin{equation}
    n = 2\nu
    \mu_0
    = T\sum_{k} \tr[\hat{G}_{k}],
\end{equation}
where $n$ is the electronic density and $\mu_0 = E_F$ is the chemical potential in the PM state.
For $c<1$, the self-consistent equation only has the trivial solution $\Delta = 0$.
For $c\geq1$, there appears another solution~\cite{Raines2024a,Raines2024b,Raines2024c}
\begin{equation}
    \Delta = c \mu_0 ,\quad \mu = (2-c)\mu_0.
    \label{eq:mf-solution}
\end{equation}
in which $\Delta$ has a finite value already at $c = 1+0^+$.
The spin-up and spin-down dispersions become
\begin{equation}
    \epsilon^{\uparrow}_k = \frac{k^2}{2m} -2 \mu_0,~~\epsilon^{\downarrow}_k = \frac{k^2}{2m} +2 \mu_0 (c-1).
    \label{eq:mf-dispersion}
\end{equation}
These expressions imply that immediately upon crossing the transition, the Fermi level in one band jumps upward to accommodate all of the electrons, while the Fermi level of the other band drops to zero.
In modern terminology, such a state is a half-metal.
We see from \cref{eq:mf-dispersion} that as the interaction strength is increased, the chemical potential of the filled band remains
fixed, while the chemical potential of the other band sinks into negative values.

It turns out, however, that the first-order nature of the transition shows up only in the ordered state, while as the transition is approached from the PM side, the static spin susceptibility diverges,
as if the transition
were continuous.
This unconventional behavior can be understood as being due to the fact that at zero temperature, the dimensionless Landau internal energy is \emph{exactly} quadratic in the spin-polarization with no higher-order terms:
~\cite{Raines2024b}
\begin{equation}
    F(\zeta) = \left(1 - c\right)\zeta^2,
    \label{eq:dimensionless-landau}
\end{equation}
where $\zeta = (n_\uparrow - n_\downarrow)/n_0$ is the spin polarization, related to $\Delta$ and $\mu$ by
\begin{align}
    \zeta(\Delta,\mu)  = {} & \frac{n_\uparrow(\Delta, \mu) - n_\downarrow(\Delta, \mu)}{n_0}                                                  \nonumber \\
    = {}                    & \frac{(\mu + \Delta)- (\mu - \Delta)\Theta(\mu - \Delta)}{ (\mu + \Delta)+ (\mu - \Delta)\Theta(\mu - \Delta) }.
\end{align}
However, once $c$ exceeds one, the slope changes sign and the equilibrium value of $\zeta$ jumps to its largest possible value $\zeta = 1$.

A first-order ferromagnetic transition has been discussed before in connection with non-analytical corrections to Fermi liquid theory~\cite{Belitz2005,Brando2016,Chubukov2004,*Maslov2009}.
In 2D, such corrections give rise to a term $-a |\zeta|^3$
with a positive $a$.
This prefactor is non-zero for a parabolic dispersion, though is small numerically.
Combining this term with \cref{eq:dimensionless-landau}, we find that the transition remains first order into a half-metal with a maximal spin polarization, but
happens at a somewhat smaller $c =1-a$.
We show in the next two sections that the ground state around a ferromagnetic QCP is a superconductor, and the gap scale does not depend critically on $c-1$. Because $a$ is numerically small, the critical $c$ still remains close to $1$ and the static uniform susceptibility increases as $1/(1-c)$ over a wide range of $(1-c)/a >1$.
As our goal is to identify the scales associated with superconductivity, we neglect the non-analytic $-a |\zeta|^3$ term in our study.

In the next two sections we consider an effective pairing interaction between low-energy fermions mediated by fluctuations of the order parameter around its equilibrium value.
In the PM phase, we will be interested in the $p-$wave pairing interaction
mediated by overdamped paramagnons.
We compute this interaction within our microscopic model and show that its $p-$wave component is weaker than previously thought based on semi-phenomenological analysis.
On the FM side, where spin-up fermions have a Fermi surface and spin-down fermions are
gapped, we first derive the interaction between spin-up and spin-down fermions, mediated by a single Goldstone magnon (a spin fluctuation
transverse to the long-range order in spin space) and a direct interaction between low-energy spin-up fermions, mediated by two Goldstone magnons.
We then derive the full spatially-odd pairing interaction between spin-up fermions by combining two single-magnon scatterings and direct two-magnon scattering.
We show that this interaction is attractive and yields a higher pairing scale than in the PM phase.

Before we proceed with the analysis, we comment on the applicability of the ladder approximation that we will be using.
It has been known for quite some time~\cite{Kanamori1963}
that this approximation over-estimates
the strength of fluctuations leading to a Stoner instability.
Moreover, numerical studies indicate that for a system with a quadratic dispersion and short-range $U$, the true ground state remains a paramagnet even when $c >1$ (see e.g., [\onlinecite{Liu2012}]).
Our reasoning to stick with the ladder approximation is four-fold.
First, there are clear experimental realizations of itinerant ferromagnetism in both 3D systems, like UGe$_2$ and ZrZn$_2$, and 2D systems, like BBG, RTG and R5G.
Quantum oscillation measurements show~\cite{Seiler2022,Zhou2022a,Zhou2021,Arp2023,Holleis2023,Zhang2023a} that a ferromagnetic state in 2D examples is a half-metal and the transition from a full to a half-metal is likely first order.
Measurements of the electronic compressibility $dn/d\mu$ show that it gets enhanced as the system approaches the FM transition~\cite{Zhou2021,Holleis2023,Patterson2024}, which can be interpreted as an indication of strong spin fluctuations on the paramagnetic side.
Both of these features (a first-order transition and strong magnetic fluctuations on the PM side) are the outcomes of the analysis of \cref{eq:S-model} in the ladder approximation.
Second, to get ferromagnetism in BBG, RTG and R5G, one would likely need to extend the model to two valleys and, possibly, also keep the momentum dependence of the gate-screened Coulomb interaction~\cite{Valenti2024,Calvera2024}
However, both analytical~\cite{Raines2024c,Calvera2024} and variational Monte-Carlo studies~\cite{Valenti2024} show that for a parabolic  dispersion, a two-valley system exhibits a direct first-order transition from a full metal to a quarter-metal with full spin and valley polarizations.
From our perspective, this implies that the system does undergo a first-order transition beyond which only a single valley is relevant and
this valley is fully spin-polarized.
Third, valley and spin orders have also been detected experimentally in ultra-clean quantum well AlAs~\cite{Shayegan2006,Gunawan2006,Hossain2020}.
This system has two valleys with elliptical fermionic dispersion in each valley.
Experiments on AlAs revealed two first-order phase transitions upon decreasing fermionic density --- the first one into a half-metal state with full valley polarization, and the second onto a quarter-metal state with full spin and valley polarizations.
Both transitions are accompanied by a rapid increase of the corresponding susceptibility.
Near a FM transition, only one valley is relevant (excitations in the other valley are all gapped), hence the system can be viewed as a single-valley one.
The way how the experimentally detected FM transition occurs then matches the outcome of the theoretical analysis within the latter approximation.
Fourth, the analysis of the pairing in the PM state, which we present below, is quite general and is rather similar to that in semi-phenomenological theories of pairing by low-energy collective modes~\cite{Klein2018,*Klein2019,Abanov2020,*Chubukov2020a}
The only distinction is in that in our microscopic theory the gradient $q^2$ term in the spin propagator appears with a small coefficient.
In the FM phase, the ladder approximation yields the correct quadratic spin-wave spectrum of transverse spin waves, and also the full interaction mediated by spin-waves obeys the Adler principle for the interaction between  fermions and Goldstone bosons.
We believe that in this respect the analysis within the ladder approximation captures the actual features of the pairing in both PM and FM phases.

\section{Paramagnetic phase: Pairing via paramagnons}
\label{sec:paramagnon-pairing}

\begin{figure}[htp]
    \centering
    \includegraphics[width=\linewidth]{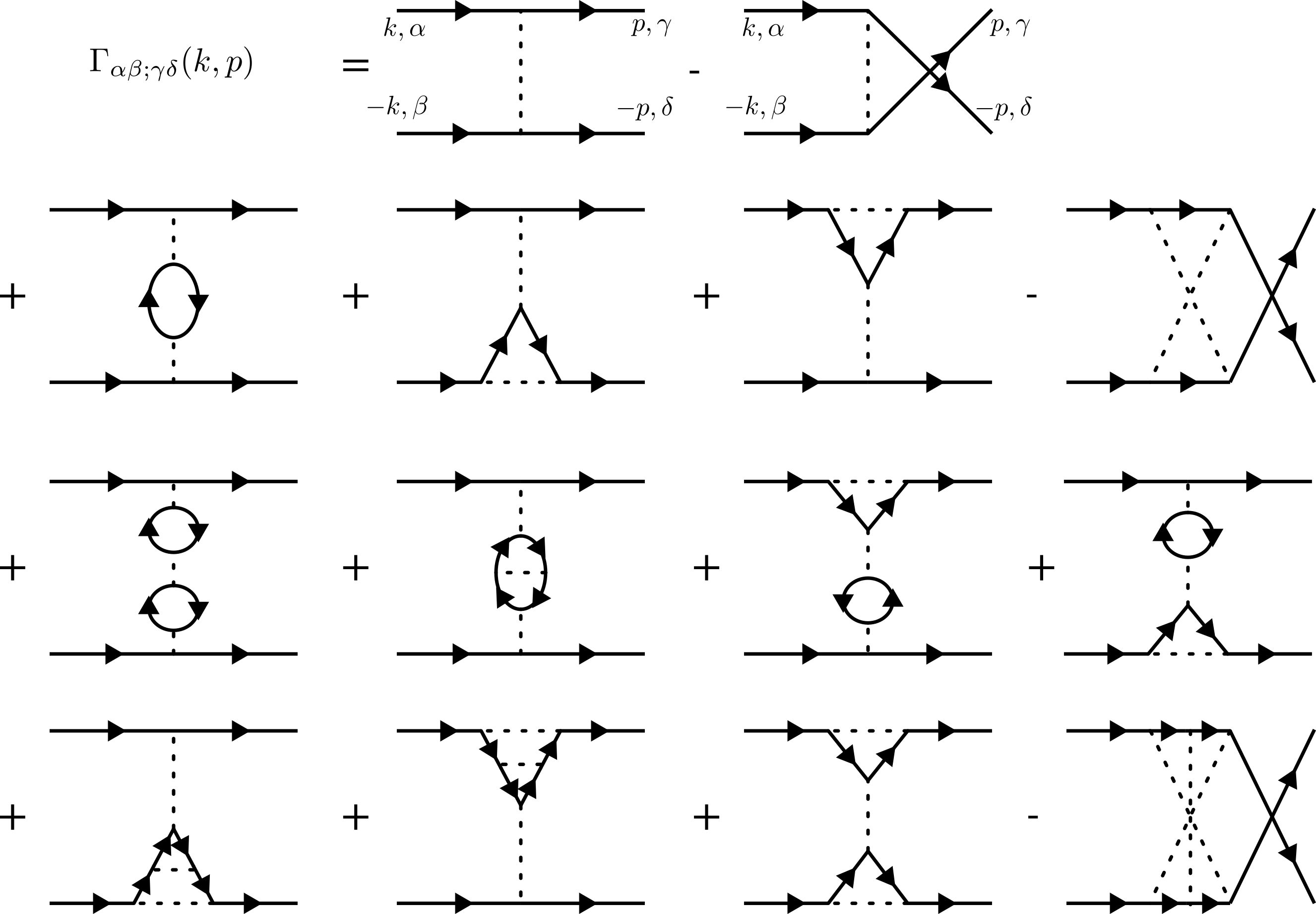}
    \caption{
        Ladder and bubble diagrams contributing to the anti-symmetrized dressed 4-fermion interaction
        $\Gamma_{\alpha \beta; \gamma \delta} (k,p)$ up to third order in $U$. Only diagrams that contain $\Pi (k-p)$ are included. }
    \label{fig:spin-interaction-vertex}
\end{figure}

We begin by considering pairing in the paramagnetic phase, $c< 1$.
The first task here is to obtain the form of the effective dynamical 4-fermion pairing interaction mediated by a paramagnon. For this, we first recall that by Pauli principle, the pairing interaction (a.k.a. pairing vertex) is the fully dressed, irreducible,  \textit{antisymmetrized}\/ interaction between fermions $\Gamma(k,p)_{\alpha \beta; \gamma \delta}$ with incoming momenta and spin projections $(\mathbf{k},\alpha)$ and $(-\mathbf{k},\beta)$  and outgoing $(\mathbf{p},\gamma)$ and $(-\mathbf{p},\delta)$ (both $\mathbf{k}$ and $\mathbf{p}$ are set to be on the Fermi surface.
To first order in $U$,
\begin{equation}
    \Gamma_{\alpha \beta; \gamma \delta} (k,p) = U\left(\delta_{\alpha \gamma} \delta_{\beta \delta} - \delta_{\alpha \delta} \delta_{\beta \gamma}\right)
\end{equation}
Using the Fierz identity, this can be re-expressed as
\begin{equation}
    \Gamma_{\alpha \beta; \gamma \delta} (k,p) = \frac{U}{2} \left(\delta_{\alpha \gamma} \delta_{\beta \delta} - \boldsymbol{\sigma}_{\alpha \beta} \cdot \boldsymbol{\sigma}_{\gamma \delta}\right)
    \label{y_1}
\end{equation}
where $\sigma^{(i)}$ are Pauli matrices.
Expressed this way, the antisymmetrized interaction contains charge and spin components, specified by $\delta_{\cdots}$ and
$\boldsymbol{\sigma}_{\cdots}$  form-factors.

The fully dressed irreducible
$\Gamma_{\alpha \beta; \gamma \delta}$ (k,p) within the ladder approximation has been discussed in earlier works~\cite{Scalapino2012,Maiti2013,Dong2023a,*Dong2023b} and we just cite the result.
The dressed $\Gamma$ can still be split into charge and spin components, but each component becomes dynamical, i.e., it depends on both the momentum transfer $\mathbf{q} = \mathbf{k} - \mathbf{p}$ and the frequency transfer $\Omega_m = \omega_{m,k}-\omega_{m,p}$.
The dependence comes via the dynamical polarization $\Pi(q, \Omega_m)$.
In explicit form
\begin{equation}
    \Gamma_{\alpha \beta; \gamma \delta} (q, \Omega_m) = \Gamma^{ch} (q, \Omega_m) \delta_{\alpha \gamma} \delta_{\beta \delta} + \Gamma^{sp} (q, \Omega_m) \boldsymbol{\sigma}_{\alpha \beta} \cdot \boldsymbol{\sigma}_{\gamma \delta}
    \label{eq:gamma-dressed}
\end{equation}
where
\begin{equation}
    \begin{aligned}
        \Gamma^{ch} (q, \Omega_m) & = & \frac{U}{2 (1+ U \Pi (q, \Omega_m))},  \\
        \Gamma^{sp} (q, \Omega_m) & = & - \frac{U}{2 (1- U \Pi (q, \Omega_m))}
        \label{eq:gamma-dressed-componenets}
    \end{aligned}
\end{equation}
and $\Pi (q, \Omega_m)$, subject to
$\Pi (0,0) = \nu$,
is the dynamical polarization bubble at momentum transfer $q$ and frequency transfer $\Omega_m$.
\Cref{eq:gamma-dressed,eq:gamma-dressed-componenets} show that near a Stoner instability at $U \nu=1$ the dominant interaction is in the spin channel, mediated by the dynamical propagator of collective ferromagnetic fluctuations.

We emphasize that although the result is intuitively expected, it is not exact even within the ladder approximation, by which we mean no mixing between bubble and  crossed diagrams, hence no contributions with the polarization bubbles with internal momenta, over which one has to integrate.
Namely, \cref{eq:gamma-dressed} is obtained by
keeping the polarization bubbles $\Pi(k-p)$ in the diagrammatic series and neglecting the bubbles  $\Pi(k+p)$.
We present the full expression for $\Gamma_{\alpha \beta; \gamma \delta}$ in \cref{app:pi}.  We show that for spin-triplet pairing, $\Pi (k+p)$ appears in the irreducible interaction in the combination $1/(1+ U \Pi (k+p))$, which does not become singular at a FM instability, while a potentially relevant  $1/(1- U \Pi (k+p))$ term cancels out.

At $T=0$, the polarization bubble can be evaluated exactly (see \cref{app:para-pi}):
\begin{equation}
    \Pi(q, \Omega_m) = \nu \left( 1- \frac{2k_F}{q}\, \Re \sqrt{ \left( \frac{q}{2k_F}+ \frac{i\Omega_m}{v_F q} \right)^2 - 1 }\right).
    \label{eq:para-pi}
\end{equation}
For the pairing we will need $q$ between the points at the Fermi surface, $q = \abs*{\mathbf{k} - \mathbf{p}} < 2k_F$.
We assume and then verify a posteriori that relevant $q$ are comparable to $k_F$, while relevant $\Omega_m$ are of order $T_c$ and are much smaller than $v_F q \sim \mu_0$.
In this situation, one can expand \cref{eq:para-pi} in $\Omega_m$.  Substituting the expansion into \cref{eq:gamma-dressed-componenets}, we obtain a spin-mediated pairing interaction of the form
\begin{equation}
    \Gamma^{sp} (q, \Omega_m) = - \frac{U}{2 \left(1-c + c \frac{|\Omega_m|}{v_F q \sqrt{1 - \left(\frac{q}{2k_F}\right)^2}}\right)}.
    \label{eq:para-gamma-sp}
\end{equation}
We see that the static interaction is independent of $q$, as long as $q <2k_F$.
This independence is a known artefact of treating $\Pi$ as a 2D static polarization bubble of free fermions.
The corrections to the bubble, which go beyond the ladder approximation,
generate a $b q^2$ term
in the denominator of \cref{eq:para-gamma-sp} (Refs. [\onlinecite{Chubukov1993,Maslov2017}])
Inserting this term into \cref{eq:para-gamma-sp} we obtain after a simple re-writing
\begin{equation}
    \Gamma^{sp} (q, \Omega_m) = - \frac{
        \bar{g}
    }{\xi^{-2} + q^2 + \gamma \frac{|\Omega_m|}
        {v_F q \sqrt{1 - \left(\frac{q}{2k_F}\right)^2}}}
    \label{eq:para-gamma-beyond-ladder}
\end{equation}
where $\bar{g} = U/(2b)$, $\gamma = \bar{g} k_F/(\pi v_F)$ and $\xi^{-2} = (1-c)/b$.
As $b$ has dimensions of $1/k^2_F$, $\bar{g}$ has dimensions of energy.
Comparing with $\Gamma^{sp} (q, \Omega_m)$ in semi-phenomenological spin-fermion theories, where the static part of the spin-mediated interaction is assumed to have the Ornstein-Zernike form $1/(\xi^{-2} +q^2)$
and the dynamical part (the Landau damping term for arbitrary $q$) is obtained from one-loop bosonic self-energies~\cite{Fay1980,Kirkpatrick2001,Kirkpatrick2003,DellAnna2006,acf,*Abanov2003,Maslov2009,*Chubukov2014,
    Fitzpatrick_15,*Wang_H_17,Lederer2015,*Schattner2016,*Lederer2017,Klein2018,*Klein2019,Abanov2020,*Chubukov2020a}
But there is one crucial difference.
In spin-fermion theories,  $\bar{g}$ is treated as a phenomenological spin-fermion vertex, which is assumed to
be smaller than the Fermi energy or, at most, comparable to it.
In our microscopic theory,
\begin{equation}
    \bar{g}  = 2 \pi \mu_0 \frac{c}{k^2_F b}.
    \label{eq:para-g-bar}
\end{equation}
For a $k^2/(2m)$ dispersion,
the value of $bk^2_F$ is fully determined by  inserting the corrections into the polarization bubble.
Such  corrections to order $U^2$ have been analyzed in~\cite{Maslov2017}
The analysis requires care as there are contributions to $b k^2_F$ from low-energy fermions and from high-energy fermions with momenta of order $q_c$, which, we remind, is the momentum cutoff for Hubbard-like interaction.
The low-energy contribution is reduced by Fermi liquid mass renormalization $\sqrt{1-c}$ (Ref. [\onlinecite{Wolfle2011}]) leaving the
high-energy contribution to $bk^2_F$ as the dominant one. This last contribution is of order one parameter-wise,
but numerically comes out as quite small for reasonable $q_c/k_F$.
We take these results as an input and set $b k^2_F$ to be a small number.
Then the effective interaction $\bar{g} \gg 2\pi \mu_0$, despite that
$U \nu \approx 1$, i.e., the actual interaction does not exceed $\mu_0$.

\begin{figure}[htp]
    \centering
    \includegraphics[width=0.8\linewidth]{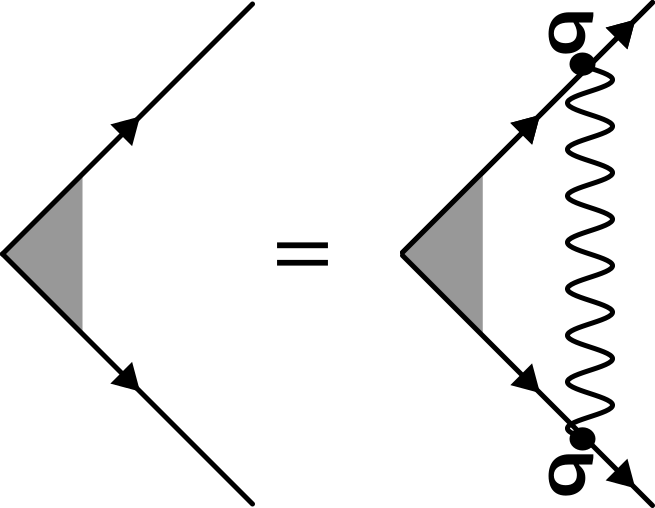}
    \caption{Linearized equation for the pairing vertex
        for paramagnon-mediated pairing.
        Solid directed lines represent fermions, and the wavy lines represent the paramagnon propagator.}
    \label{fig:pairing-vertex}
\end{figure}

At a first glance, a larger effective spin-fermion coupling ${\bar g}$ should boost superconducting $T_c$. We show, however, that this is not the case as in this situation the $p-$wave component of the interaction actually decreases.
To see this, we now associate $\Gamma^{sp} (q, \Omega_m)$, given by \cref{eq:para-gamma-beyond-ladder}, with the spin-mediated pairing interaction,  and
analyze the  linearized equation for the pairing vertex, $\hat{\phi}_{k}$, where here we set $k = ({\bf k}, \omega_{m,k})$. The temperature at which this equation has a non-trivial solution is a superconducting $T_c$. We assume that $T_c$ is small and use the $T=0$ form of
$\Gamma^{sp} (q, \Omega_m)$. The equation for $\hat{\phi}_{k}$ is shown graphically in  \cref{fig:pairing-vertex}.
In analytic form,
\begin{equation}
    \hat{\phi}_{k} =
    -
    T\sum_{k'i}G_{k'}G_{-k'}\Gamma^{sp}(k-k')\hat{\sigma}^{i}\hat{\phi}_{k'}(\hat{\sigma}^{i})^{T}
    \label{eq:lin-pairing}
\end{equation}
where $G_{k}$ is the Matsubara fermion Green's function in the normal state and $\Gamma^{sP} (k-k')$ is given by \cref{eq:para-gamma-beyond-ladder}.
Decomposing the pairing vertex into singlet and triplet channels,
\begin{equation}
    \hat{\phi}_{k} = i\hat{\sigma}^{2}\left(\phi^{0}_{k} + \mathbf{d}_{k}\cdot \bm{\sigma}\right)
\end{equation}
and using
\begin{equation}
    \sum_{i} \sigma^{i}\sigma^{y}(\sigma^{i})^{T} = -3\sigma^{y},\
    \sum_{i} \sigma^{i}i\sigma^{j}\sigma^{y}(\sigma^{i})^{T} = i\sigma^{j}\sigma^{y}.
\end{equation}
we find that
the interaction mediated by ferromagnetic spin fluctuations is repulsive in the singlet channel and attractive in the spin triplet channel.
Without loss of generality, we then choose
$\hat{\phi}_{k} = \phi_{n} f(\mathbf{k})\hat{\sigma}^{0}$, where $f(\mathbf{k})$ is
odd in momentum and $\phi_{n} = \phi_{-n}$.
The linearized gap equation, \cref{eq:lin-pairing}, then reduces to
\begin{equation}
    \hat{\phi}_{n}f(\mathbf{k}) =
    -
    T\sum_{k'}G_{k'}G_{-k'}\Gamma^{sp}(k-k')\phi_{n'}f(\mathbf{k'}).
    \label{eq:para-lin-pairing}
\end{equation}
Substituting the forms of the free fermion propagators, keeping the $|\mathbf{k}'|$ dependence only in the fermionic propagators and setting $|\mathbf{k}|=|\mathbf{k}'|=k_{F}$ elsewhere, as is usually done in the Eliashberg-like treatment, and integrating over the dispersion of an intermediate fermion,
we re-express \cref{eq:para-lin-pairing} as an integral equation over the angle on the Fermi surface and Matsubara frequency:
\begin{align}
    \phi_{n}f(\theta)
    = {} &
    -
    \nu\pi T\sum_{n'} \frac{\phi_{n'}}{|\omega_{n'}| }\oint \frac{d\theta'}{2\pi}f(\theta') \nonumber \\
         & \times
    \Gamma^{sp} \left(\omega_{n}-\omega_{n'}, 2 k_{F}\sin\left|\frac{\theta-\theta'}{2}\right|\right),
    \label{eq:para-factorized}
\end{align}
where  $\omega_n = \pi T (2n+1)$, $\theta$  and $\theta'$ are the angles between the directions of $\mathbf{k}$ and $\mathbf{k}'$ on the Fermi surface and
$\abs*{\mathbf{k} - \mathbf{k'}} = 2k_F \sin{(\theta-\theta')/2}$.
For the electron-phonon problem and many quantum-critical pairing problems, the factorization of the momentum integration can be justified by the condition that a pairing boson is a slow mode compared to an electron~\cite{Chubukov2020b}
Here, the issue is less clear, but for an order of magnitude estimate of the pairing energy scale using the Eliashberg approximation should not be problematic.

Going forward, we focus on the $p$-wave ($\ell =1$) component of $f(\theta)$ ($f(\theta)=\cos\theta$) for which
generically the instability is the strongest.
Multiplying both sides of the gap equation by $\cos\theta$ and integrating over the Fermi surface we obtain the effective gap equation as an integral equation in frequency only:
\begin{equation}
    \phi_{n}  =
    - \pi T \sum_{n'} \tilde{\Gamma}^{sp}_{\ell=1} (\omega_{n} - \omega_{n'}) \frac{\phi_{n'}}{|\omega_{n'}|}
    \label{eq:para-gap-equation}
\end{equation}
where
\begin{multline}
    \tilde{\Gamma}^{sp}_{\ell=1}(\Omega_{m}) = \nu\int_{0}^{\pi/2} \frac{d\theta}{\pi} \cos \theta \\
    \times \left[ \Gamma^{sp} (\Omega_{m}, 2k_{F}\sin\frac{\theta}{2}) - \Gamma^{sp}(\Omega_{m}, 2k_{F}\cos\frac{\theta}{2}) \right].
    \label{eq:para-dimensionless-vertex}
\end{multline}
Writing the interaction in this way emphasizes that the triplet channel sees the anti-symmetrized interaction $\Gamma^{sp}(\mathbf{k} - \mathbf{p}) - \Gamma^{sp}(\mathbf{k}+\mathbf{p})$.

Up to this moment, the analysis was not specific to the value of $\bar{g}/\mu_0$, which, we recall, depends on the magnitude of $b k^2_F$, \cref{eq:para-g-bar}.
In previous studies of \cref{eq:para-gap-equation}, with a semi-phenomenological form of $\Gamma^{sp} (\Omega_{m}, k_{F}\sin\frac{\theta}{2})$, it was assumed that $b \sim (1/mW)$ such that  $b k^2_F >1$ and $\bar{g}/(2\pi) \mu_0$ is at most of order one.
In this situation, relevant momentum transfers $q = 2k_F \sin{\theta}/2$ are small in $1/(b k^2_F)$, as we will see immediately below, and out of the two terms in \cref{eq:para-gap-equation} only the first one is relevant.
Keeping only this term in \cref{eq:para-gap-equation}, approximating $\sin(\theta/2)$ by $\theta/2$ and $\sin{\theta}$ by $\theta$, integrating in \cref{eq:para-gap-equation} over $\theta$ and setting $c=1$, we obtain \cref{eq:para-gap-equation} in the form
\begin{equation}
    \phi_{n} = 0.039 \left(\frac{\bar{g}^2}{T \mu_0}\right)^{1/3} \sum_{n'} \frac{\phi_{n'}}{|2n'+1| |n-n'|^{1/3}}.
    \label{eq:para-gap-equation-simplified}
\end{equation}
From a general point of view this describes quantum-critical pairing wit the exponent $\gamma =1/3$ (Refs.~[\onlinecite{Abanov2020,*Chubukov2020a}]).
The formally divergent $n=n'$ term in the right hand side accounts for the pair-breaking effect from thermal fluctuations.
As our goal is to determine the characteristic pairing scale at $T=0$, where thermal fluctuations are not present, we drop this term and associate the pairing scale with $T^*$, at which \cref{eq:para-gap-equation-simplified} has at solution.
It is clear from \cref{eq:para-gap-equation-simplified} that $T^* \sim \mu_0 (\bar{g}/(2\pi \mu_0))^2$.
Solving \cref{eq:para-gap-equation-simplified} numerically, we find
\begin{equation}
    T^* = 0.065 \mu_0 \left(\frac {\bar g}{2\pi \mu_0}\right)^2.
    \label{eq:T-para-large-bkf}
\end{equation}
A more accurate result for the numerical prefactor ($0.022$ instead of $0.065$) is obtained by
including fermionic  self-energy $\Sigma(\omega_m)$ into the calculation of  $T^*$ (Refs. \cite{Klein2018,*Klein2019,Abanov2020,*Chubukov2020a,Zhang2024}).
However, the functional form of $T^*$ remains intact because
$\bar{g}^2/\mu_0$ is the energy scale at which $\Sigma (\omega_m) \sim \omega_m$.
Returning to the integral over $\theta$ and using $2\pi T^*$ as a proxy for $\Omega_m$, we estimate typical $\theta$ as
$O(\bar{g}/(2\pi \mu_0))$.
We see that typical $\theta$ are small when $\bar{g}/(2\pi \mu_0)$ is small.

In our case, $b k^2_F$ is a small number, and $\bar{g}/(2\pi E_F)$ is a large number.
In this situation, we cannot assume that $\theta$ is small and then have to keep both terms in
$\tilde{\Gamma}^{sp}$ in \cref{eq:para-gap-equation}.
Substituting the full form of
$\tilde{\Gamma}^{sp}$ into the gap equation, we obtain
\begin{equation}
    \phi_n = \frac{1}{8} \frac{\bar g}{2\pi \mu_0} \pi T \sum_{n'} \Psi
    \left(|\omega_n-\omega_{n'}|\right)  \frac{\phi_{n'}}{|\omega_{n'}|}
    \label{eq:gap-function-C}
\end{equation}
where
$\Psi (\Omega_m)$, proportional to $\tilde{\Gamma}^{sp} (\Omega_m)$, is
\begin{equation}
    \Psi (\Omega_m) \equiv \int_0^{\pi/2} \frac{d \theta}{\pi} \frac{\sin^2{\theta} \cos^2{\theta}}{(\sin{\theta} \sin^2{\theta\over2} + u)(\sin{\theta} \cos^2{\theta\over2} + u)}
    \label{eq:C-def}
\end{equation}
where $u = \left(\pi {\bar g}|\Omega_m|/(4(2\pi \mu_0)^2)\right)$.
We assume and justify post-hoc that for large $\bar{g}/(2\pi \mu_0)$,  $T_c$ is such that for $\Omega_m \sim T_c$,
$u$ is a large number. Then $\Psi (\omega_m) =(2\pi \mu_0)^4/(\pi^2 {\bar g}^2 \Omega^2_m)$.  Substituting into \cref{eq:gap-function-C}, we re-express the gap equation as
\begin{equation}
    \phi_n = \frac{1}{32 \pi^4} \frac{(2\pi \mu_0)^3}{{\bar g} T^2} \sum_{n'} \frac{\phi_{n'}}{|2n'+1| |n-n'|^2}
    \label{eq:gap-equation-large-u}
\end{equation}
From a general perspective, this corresponds to quantum-critical pairing with the exponent $\gamma =2$ (Refs.~[\onlinecite{Abanov2020,*Chubukov2020a}]).
From dimensional analysis, $T^* \sim(2\pi \mu_0)^{3/2}/{\bar g}^{1/2}$.  Solving \cref{eq:gap-equation-large-u} numerically, again dropping the $n=n'$ term, we obtain
\begin{equation}
    T^* = 0.13 \mu_0 \left(\frac{2\pi \mu_0}{{\bar g}}\right)^{1/2}.
    \label{eq:T-para-small-bkf}
\end{equation}
Substituting
$2\pi T^*$ as a proxy for $\Omega_m$ into $u = \pi \bar{g} \Omega_m/(4 (2\pi \mu_0)^2)$, we obtain $u = 0.1
    (\bar{g}/(2\pi \mu_0))^{1/2}$.
We see that $u$ is indeed large when ${\bar g}/(2\pi \mu_0)$ is a sufficiently large number.

We see from \cref{eq:T-para-small-bkf} that at large spin-fermion coupling $\bar{g}$, the characteristic pairing scale decreases
as $1/\sqrt{\bar{g}}$, contrary to a naive expectation that a larger interaction should give rise to a larger pairing scale.
Digging into the gap equation, we see that the reason is that in a situation when the Landau damping term in $\Gamma^{sp}$ becomes the dominant one, the $p$-wave component of the pairing interaction drops as the Landau-damping term is invariant under $\theta \to \pi-\theta$, which changes the sign of the $p$-wave form factor.

The two expressions for $T^*$, \cref{eq:T-para-small-bkf} and \cref{eq:T-para-small-bkf},
can be combined to the scaling formula
\begin{equation}
    T^* = 0.13 \mu_0 \left(b k^2_F\right)^{1/2}  \Phi (b k^2_F)
    \label{eq:T-star-scaling}
\end{equation}
where $\Phi (0) =1$ and $\Phi (x \gg 1) = 1/(2x^{5/2})$.  We recall that \cref{eq:T-para-small-bkf} is the result of previous, semi-phenomenological studies, which assumed that the $b q^2$ term in the static $\Gamma^{sp}$ is large enough such that $b k^2_F >1$, while \cref{eq:T-star-scaling} is the result of our analysis of the microscopic model with a parabolic dispersion, for which $b k^2_F$ is small.

\begin{figure}
    \centering
    \includegraphics[width=\linewidth]{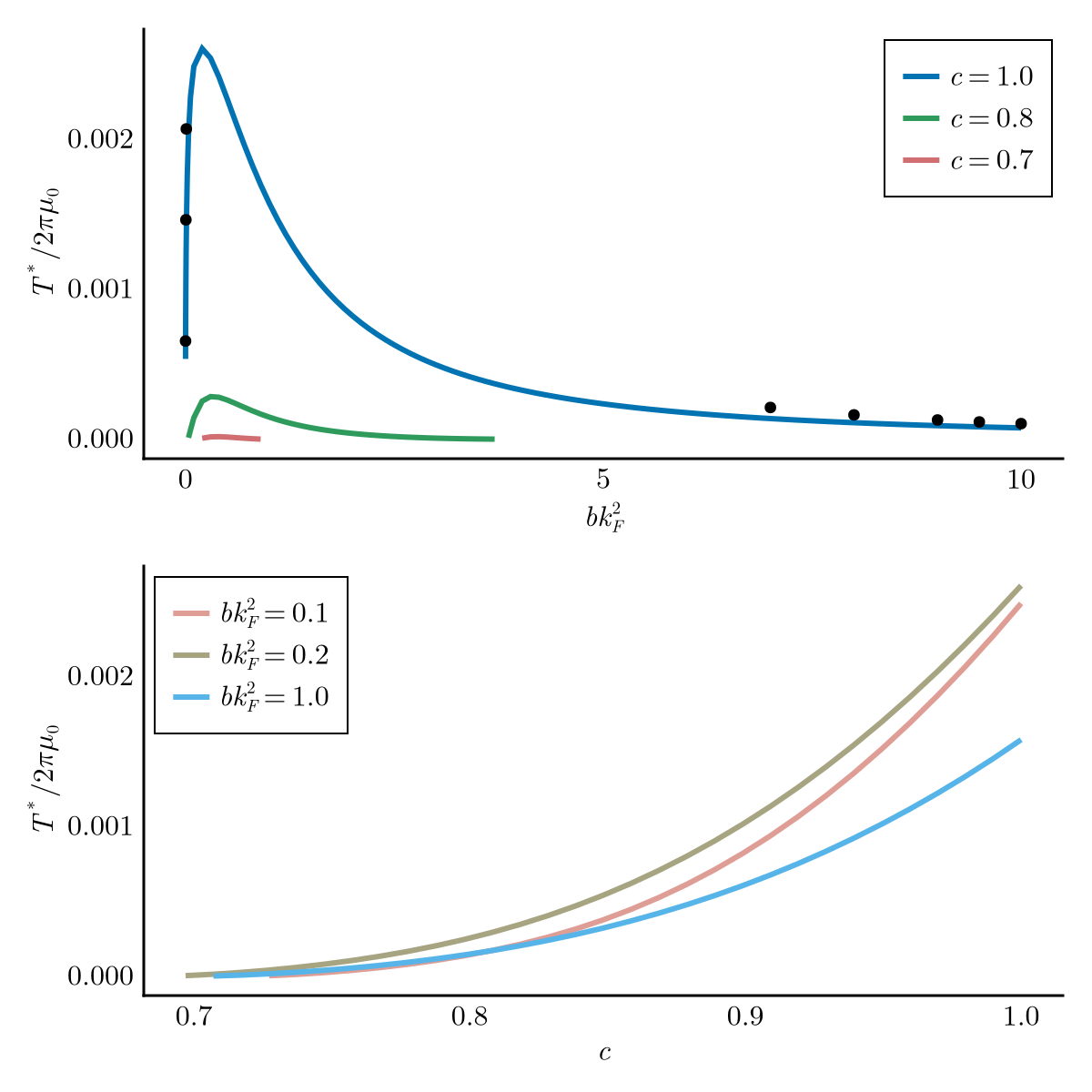}
    \caption{(Color Online) The pairing scale $T^*$  for
        paramagnon-mediated triplet pairing
        near a FM QCP, as a function of the prefactor $b$ for the $q^2$ term in the bosonic propagator and deviations from a FM QCP,
        measured by $1-c$.
        Dots on the upper plots represent the analytical results \cref{eq:T-para-small-bkf} and \cref{eq:T-para-large-bkf} for the asymptotic behavior of $T^*$ at small and large $b k^2_F$, respectively.
        \label{fig:tpnormal}}
\end{figure}

In \cref{fig:tpnormal}
we present the results of the numerical solution of the gap equation for different $b k^2_F$ both at the FM QCP ($c=1$), and away form it ($c <1$).
For numerical calculation, we defined $u_n = \frac{\phi_{n}}{\sqrt{|\omega_{n}|}}$ in order to make the kernel
of the gap equation \cref{eq:para-gap-equation} symmetric, and truncated the sum over Matsubara frequencies at a maximum $n_\text{max}=50$
\begin{equation}
    u_n =
    -  \sum_{n'=-n_\text{max}}^{n_\text{max}} \frac{\pi T\tilde{\Gamma}^{sp}_{\ell=1} (n-n')}{\sqrt{|\omega_{n}\omega_{n'}|}}u_{n'}\pod{n' \neq n}.
\end{equation}
We identify $T^*$ with the temperature at which the largest eigenvalue of the $(2n_\text{max}+1)\times (2n_\text{max}+1)$ kernel matrix
\begin{equation}
    K_{nn'} =
    \begin{cases}
        -
        \frac{\pi T {\tilde \Gamma}^{sp}_{\ell=1} (n-n')}{\sqrt{|\omega_{n}\omega_{n'}|}} & n' \neq n \\
        0                                                                                 & n' = n
    \end{cases}
\end{equation}
is equal to 1.

In the top panel of \cref{fig:tpnormal} we plot
$T^*/(2\pi \mu_0)$ as a function of $b k^2_F = (2\pi \mu_0)/{\bar g}$ for different $c$.
We clearly see that at small $b k^2_F$,  $T^*$ scales as $(b k^2_F)^{1/2}$.
The square-root dependence \cref{eq:T-para-small-bkf} is overlaid on the numerical results in the top panel of \cref{fig:tpnormal}, for small $b k^2_F$ and shows good qualitative agreement.
At larger $b k^2_F$, $T_c$ passes through a maximum and then decreases as $1/(b k^2_F)^2$.
We emphasize that this last behavior, which, we remind, was found in previous studies assuming that relevant momentum transfers are small,  holds only for numerically large $b k^2_F$.
For realistic $b k^2_F = O(1)$, the system is in the crossover region between the two regimes.
In the lower panel, we plot
$T^*/(2\pi \mu_0)$  as a function of $c$ for different $b$.  We clearly see that
the magnitude of $T^*$ drops away from $c=1$.

\begin{figure}
    \centering
    \includegraphics[width=\linewidth]{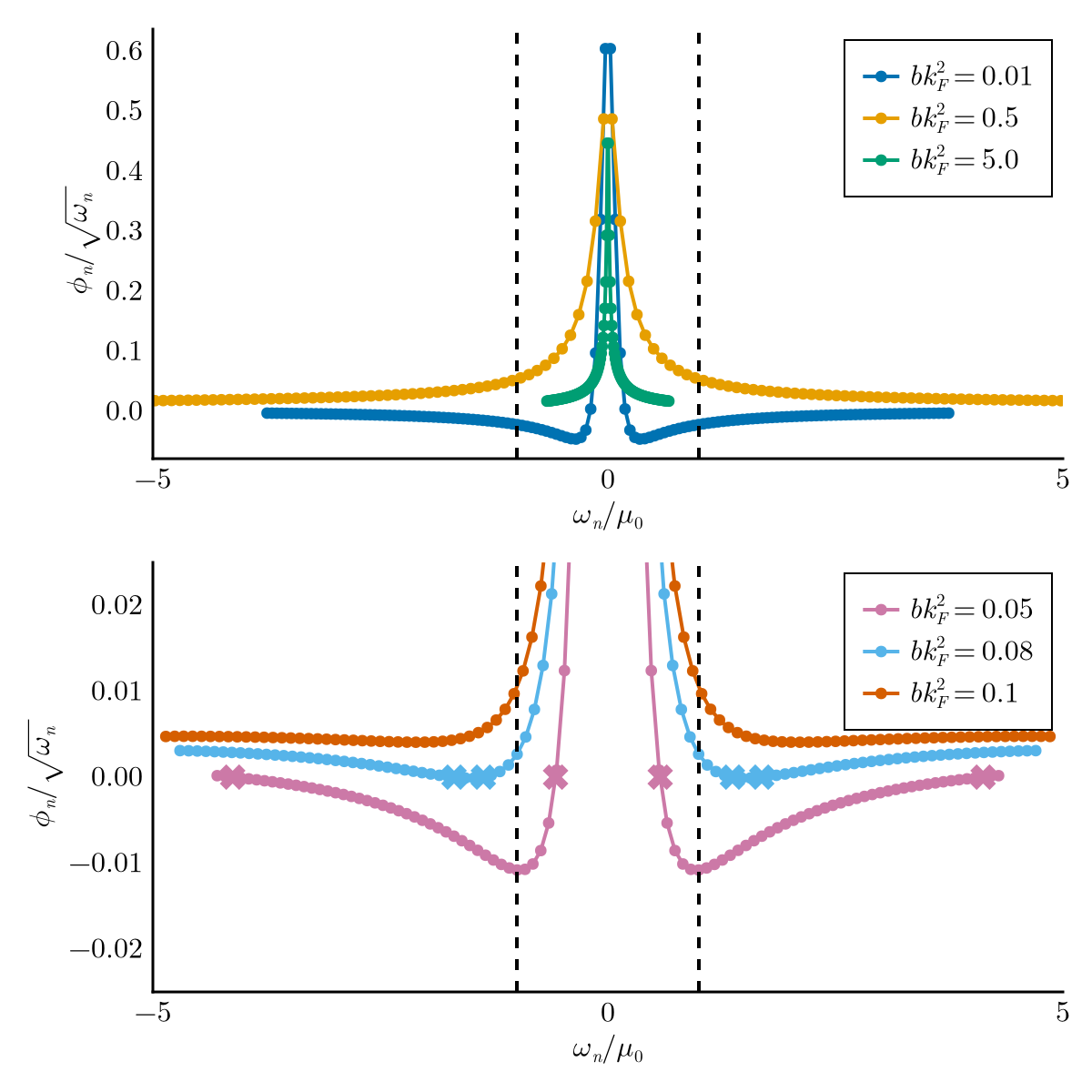}
    \caption{
        Top: Eigenfunction of the kernel matrix $K_{nn'}$ with maximal eigenvalue for different $b k^2_F$ and $c$.
        For small or large $b k^2_F$, the eigenfunctions
        $\phi_n$ are mostly confined to low frequencies,
        justifying the asymptotic forms of the polarization bubble $\Pi(q, \Omega_m)$ in \cref{eq:para-gap-equation}.
        For intermediate $b k^2_F$, the eigenfunctions have significant contributions from high frequencies $\omega_n \sim \mu_0$,
        where the expansion in $\omega_n$ does not hold and the calculation of $T^*$ has to be done numerically.
        Bottom: Details of the frequency  dependence of the eigenfunction for the maximal eigenvalue.
        At small,
        $bk^2_F$, the eigenfunction changes sign twice on the Matsubara axis.
        The first sign change is indicated by the cross marks.
        As $b k^2_F$ increases, the
        position of two sign changes move closer together until they collide and move away from the Matsubara axis (see \cref{app:b=0}).
        \label{fig:eigenfunctions}}
\end{figure}

Finally, we note that there is a qualitative change between superconductivity at small and large $bk^2_F$.  We see from the bottom panel of \cref{fig:eigenfunctions} that at small $b k^2_F$, the gap function $\phi_n$, changes sign twice on the Matsubara axis.
On a more careful look we found that this behavior originates from the fact that at $b=0$, the dynamical pairing interaction vanishes in the $p-$wave channel, if we use \cref{eq:para-gamma-sp}, but
turns out to be repulsive if we do not expand in frequency (see \ref{app:b=0} for details).
This repulsion competes with the attraction induced by the $b q^2$ term in the spin propagator in \cref{eq:para-gamma-beyond-ladder} and for small $bk^2_F$ accounts for the sign changes of $\phi_n$

A zero of $\phi_n$ on the Matsubara axis is the origin of a dynamical vortex (the phase of the gap function changes by $2\pi$ under anti-clockwise rotation around this point), Refs.~[\onlinecite{Chubukov2019,*Pimenov2022a,Christensen2021}]
A dynamical vortex
in the upper  half-plane of frequency
affects the behavior of the complex gap function along the physical real frequency axis, $\phi (\omega) = |\phi (\omega)| e^{i \psi (\omega)}$
--- it gives rise to a $2\pi$ phase slip between $\omega = - \infty$ and $+\infty$ (or, equivalently, by $\pi$ between $\omega =0$ and $\omega = \infty$). The two vortices  in the upper half-plane give rise to $4\pi$ phase slip on the real axis.
As $b k^2_F$ increases, these two vortices come closer to each other, merge, and then disperse in  opposite directions away from the Matsubara axis, bending towards the real axis.   They eventually cross the real axis at $b = b_{c}$ and  continue dispersing into the lower  half-plane (Ref.~\onlinecite{Christensen2021}).
Once the vortices leave the upper half-plane, they no longer cause a $4\pi$ phase slip.
From this perspective,  $b_{c}$ can be thought of as a
point of a topological transition from
a topologically nontrivial pairing state with dynamical vortices in the upper half-plane of frequency causing
$4\pi$ phase slip on the real frequency axis, to a topologically trivial pairing state with no vortices.

\section{Ferromagnetic phase: Pairing via Goldstone modes}
\label{sec:magnon-pairing}
\begin{figure}
    \includegraphics[width=0.8\linewidth]{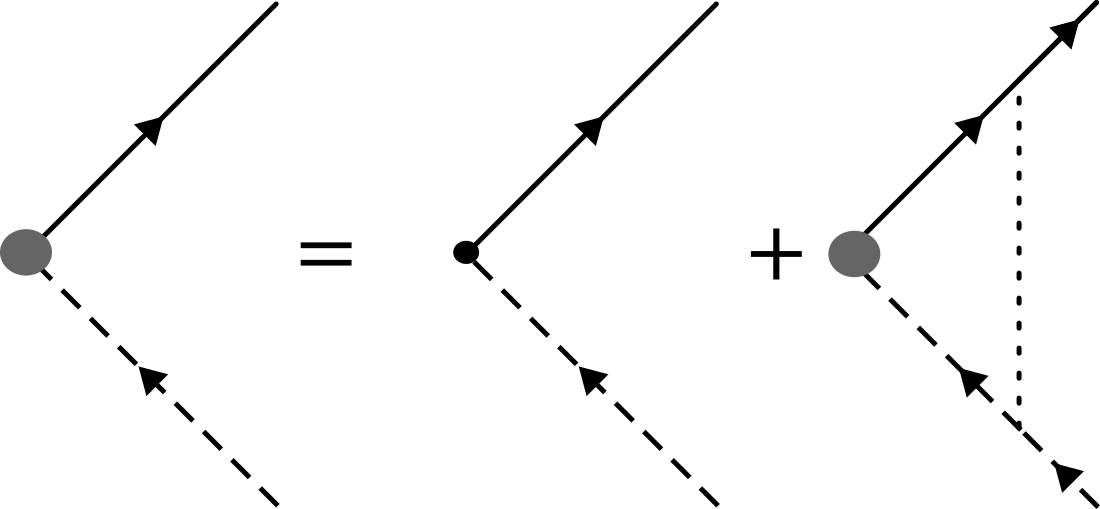}
    \caption{Bethe-Salpeter
        equation
        for the renormalized transverse spin vertex $\hat{\sigma}_+$.
        Solid
        lines are
        for fermions in
        the occupied (spin-up) band, and dashed
        lines are
        for fermions
        in the unoccupied (spin-down) band.
        The undirected dashed line is the bare Hubbard interaction $U$.
        The equation for $\hat{\sigma}_-$ is obtained by reversing the fermion lines.
        \label{fig:spin_vertex}}
\end{figure}
We now derive the pairing interaction in the ferromagnetically ordered state.  We begin by deriving the magnon propagator and the fermion-magnon vertex.
For this we note that for order along $z$-direction, Goldstone magnons are the poles of transverse susceptibilities $\chi_{xx}$ and $\chi_{yy}$.
Both the $\hat{\sigma}_x$ and $\hat{\sigma}_y$ vertices
contain combinations $\bar{\psi}^\uparrow_{k +q, \omega_n+\Omega_m} \psi^\downarrow_{k, \omega_n}$ and  $\bar{\psi}^\downarrow_{k +q, \omega_n+\Omega_m} \psi^\uparrow_{k, \omega_n}$.
To get magnons, we then introduce two trial
vertices $\hat{\sigma}_x$ and $\hat{\sigma}_y$ (or, equivalently, $\hat{\sigma}_{+}$ and $\hat{\sigma}_{-}$,
as
shown in \cref{fig:spin_vertex},
and convert each vertex into a fully dressed one by including an infinite series of renormalizations set by $U$.
To be consistent with the analysis in \cref{sec:model}, we only keep ladder renormalizations.
The ladder summation is straightforward and yields the
susceptibility $\chi_{\uparrow \downarrow} (q, \Omega_m)$ - the ratio of the fully dressed and the bare vertices,  in the form
\begin{equation}
    \chi_{\uparrow \downarrow} (q, \Omega_m) = \dfrac{1}{1 - U \Pi_{\uparrow \downarrow} (q, \Omega_m)},
\end{equation}
where
\begin{equation}
    \Pi_{\uparrow \downarrow} (q, \Omega_m) = - \int \frac{d^2k}{4\pi^2} \int \frac{d\omega_n}{2\pi}
    G^{\uparrow} (k+q, \omega_n + \Omega_m) G^{\downarrow} (k, \omega_n).
    \label{eq:Pi_up_down}
\end{equation}
The convention of notations is such that the outgoing fermion has spin-up, and $\Omega_m$ and $q$ are incoming
frequency and momentum.
Similarly,
\begin{equation}
    \chi_{\downarrow \uparrow} (q, \Omega_m) = \frac{1}{1 - U \Pi_{\downarrow \uparrow} (q, \Omega_m)}
\end{equation}
where again $\Omega_m$ and $q$ are incoming momenta.
Evaluating $\Pi$ and expanding in $\Omega_m$ and $q$, we obtain
\begin{equation}
    \begin{gathered}
        \chi_{\uparrow \downarrow} (q, \Omega_m) = \frac{2\mu_0 c}{i \Omega_m + \frac{q^2}{2m} \frac{c-1}{c}} \\
        \chi_{\downarrow \uparrow} (q, \Omega_m) = \frac{2\mu_0 c}{-i \Omega_m + \frac{q^2}{2m} \frac{c-1}{c}}.
    \end{gathered}
\end{equation}
We see that $\chi_{\downarrow \uparrow} (q, -\Omega_m) = \chi_{\uparrow \downarrow} (q, \Omega_m)$.
This will be relevant to our analysis below.

The physical $\chi_{xx}$ and $\chi_{yy}$ are linear combinations of  $\chi_{\uparrow \downarrow} (q, \Omega_m)$ and
$\chi_{\downarrow \uparrow } (q, \Omega_m)$.
Each contains two poles corresponding to two magnon modes running in opposite directions. Both poles are located in the lower half-plane of complex frequency, infinitesimally close to the real axis.

\begin{figure}
    \includegraphics[width=0.6\linewidth]{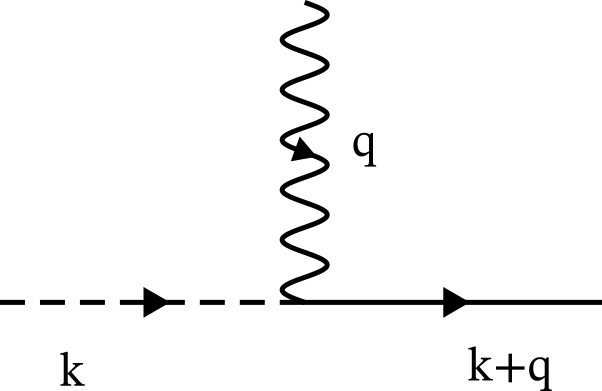}
    \caption{
        Fermion-magnon
        vertex.
        The solid and dashed lines denote fermions from the
        occupied (spin-up) band,
        the unoccupied (spin-down) band,  respectively, and the wavy line is the magnon propagator.
        Note the vertex involves one spin-up and one spin-down fermion.
        \label{fig:fermion-magnon-vertex}}
\end{figure}
\begin{figure}
    \includegraphics[width=0.6\linewidth]{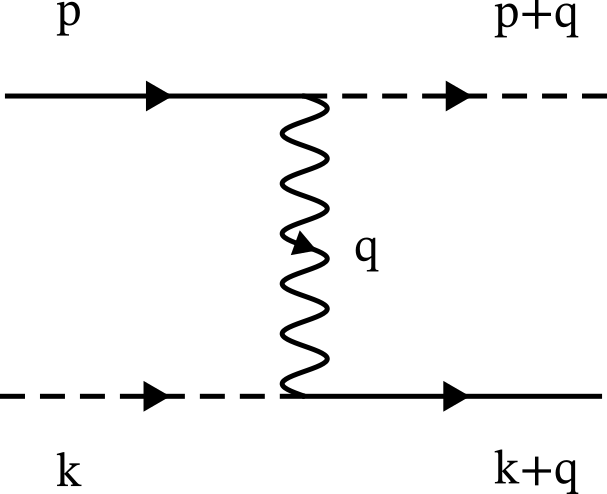}
    \caption{Effective 4-fermion interaction mediated by a single magnon.
        The solid fermion lines are in the occupied (spin-up) band, and the
        dashed ones are in the unoccupied (spin-down) band.
        The wavy line is the magnon propagator.
        The effective interaction has one outgoing and one incoming spin-up fermion, and one outgoing and one incoming spin-down fermion.
        \label{fig:effective-interaction}}
\end{figure}

Our next goal is to obtain an effective 4-fermion interaction mediated by Goldstone bosons.
For this we notice that (i) a fermion-magnon vertex  is between
a spin-up and spin-down fermion (see \cref{fig:fermion-magnon-vertex}) and (ii) a magnon-mediated
pairing interaction
must contain one incoming and one outgoing spin-up fermion and one incoming and one outgoing spin-down fermion (see \cref{fig:effective-interaction}).
There is no magnon-mediated
pairing interaction with two incoming spin-up fermions and two outgoing spin-down fermions and vice versa.

\begin{figure}
    \includegraphics[width=0.9\linewidth]{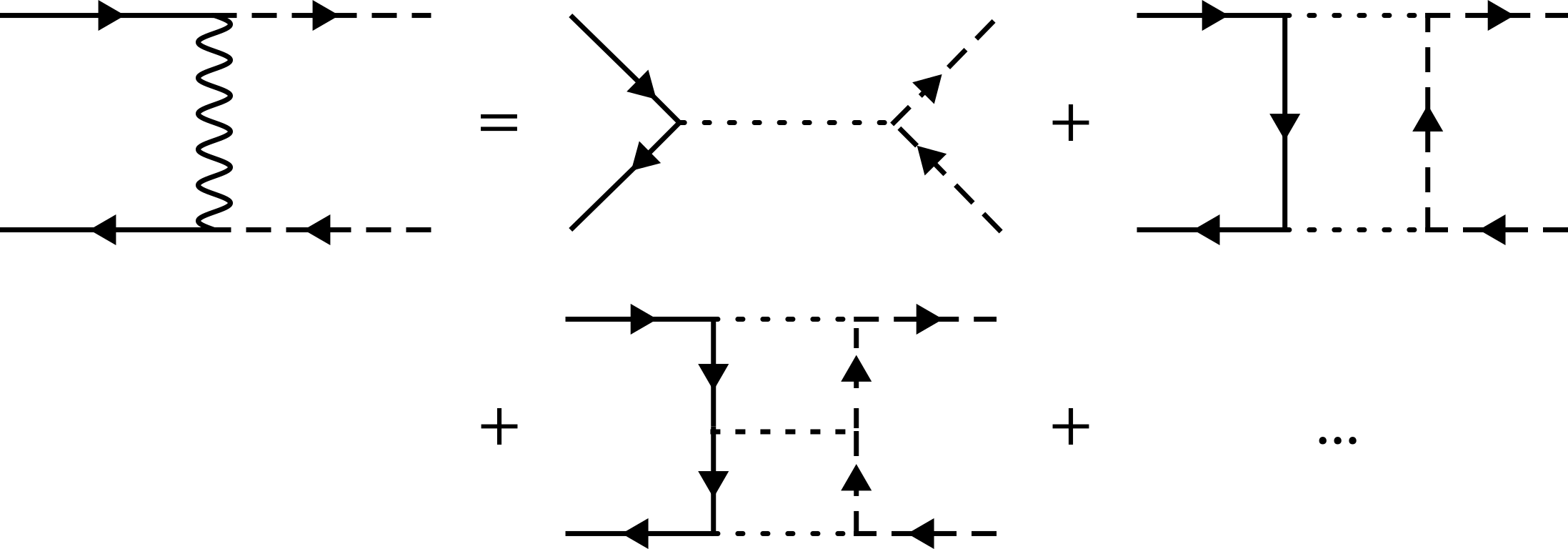}
    \caption{Diagrammatic representation of the magnon-mediated 4-fermion interaction in the ferromagnetic state.
        The solid fermion lines represent gapless (spin-up) fermions, and the
        dashed fermion lines represent the gapped (spin-down) fermions.
        The wavy lines represent the magnon propagator, and the undirected dotted line is the Hubbard interaction $U$.
        \label{fig:paramagnon-bs-fm}}
\end{figure}
In \cref{fig:paramagnon-bs-fm} we show diagrammatically how to obtain a magnon-mediated 4-fermion interaction starting from the Hubbard $U$
between the densities of spin-up and spin-down fermions and collecting ladder series of diagrams  with $\Pi_{\uparrow \downarrow} (q, \Omega_m)$ in each cross-section.
The ladder summation is straightforward and yields the magnon-mediated interaction
in the form \begin{equation}
    \Gamma_{1} (q, \Omega_m) = U  \chi_{\uparrow \downarrow} (q, \Omega_m)= U \frac{2\mu_0 c}{i\Omega_m + \frac{q^2}{2m} \frac{c-1}{c}}.
    \label{eq:gamma_1_definition}
\end{equation}
This specific form is for the fully spin polarized state, where at the same momentum the energy of a spin-up fermion is $\epsilon^{\uparrow}_k = k^2/(2m) - 2\mu_0$, while for a spin-down fermion it is $\epsilon^{\downarrow}_k = k^2/(2m) + 2\mu_0 (c-1)$.
Notice that the prefactor in \cref{eq:gamma_1_definition} does not vanish at $q = \Omega_m =0$
for the magnon pole structure is $q-$independent.
A similar result has been obtained recently in Ref.~\onlinecite{Dong2024}.
According to [\onlinecite{Dong2024}],
the interaction in \cref{fig:paramagnon-bs-fm} should be viewed as a non-diagonal term in the $2\times 2$ basis.
We note in passing that this result is specific to an ordered ferromagnet, in which spin-up and spin-down excitations are split.
For an ordered antiferromagnet, the situation is different because excitations in the ordered state remain spin-degenerate, and there is a direct magnon-mediated interaction between low-energy fermions.
The side vertex for such
interaction contains an overall factor of $q$,  in agreement with
the Adler principle for Goldstone bosons interacting with low-energy fermions~\cite{Schrieffer1989,Schrieffer1995,Sokol1995,*Morr1997,*Morr_1997_a,Sushkov,*Sushkov_1,Ismer_2010}.
\begin{figure}
    \includegraphics[width=0.8\linewidth]{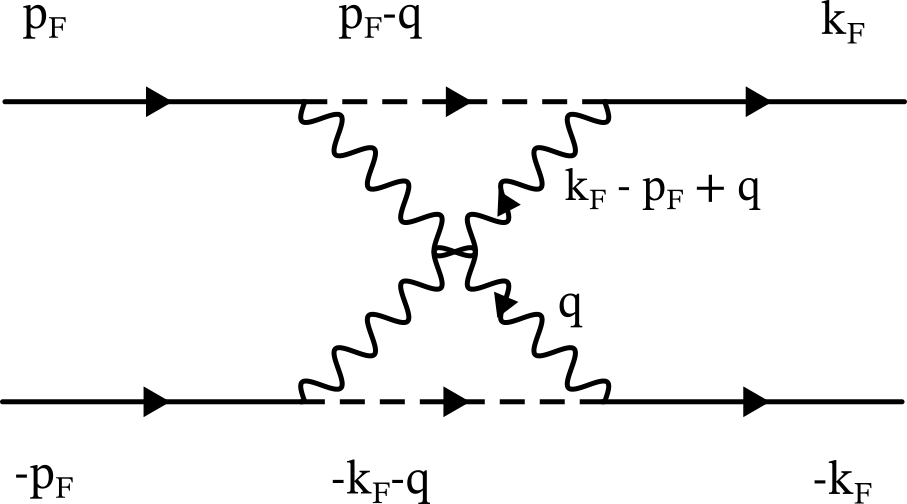}
    \caption{Effective magnon-mediated interaction between two spin-up fermions.
        The solid
        lines are fermionic propagators  in the occupied (spin-up) band, and the dashed
        lines
        are
        fermionic propagators
        in the unoccupied (spin-down) band.
        The wavy line is the magnon propagator.
        \label{fig:effective-2magnon}}
\end{figure}

To obtain the effective interaction with only spin-up fermions with momenta on the Fermi surface, we need to
keep the magnon-mediated interaction to second order.
The corresponding diagram is shown in \cref{fig:effective-2magnon}.
It involves the convolution of the two magnon propagators.
Because we assumed that the magnon momentum $q$ is small, fermionic $\mathbf{k}_F$ and $\mathbf{p}_F$ have to be close, i.e., $\delta\mathbf{k} = \mathbf{k}_F - \mathbf{p}_F$  has to be small.
We label this interaction
as $\Gamma_{2a} (\mathbf{k}_F - \mathbf{p}_F) = \Gamma_{2a} (\delta\mathbf{k})$.
In explicit form,
\begin{widetext}
\begin{align}
    \Gamma_{2a} (\delta\mathbf{k}) = {} & - U^2 \int\frac{d^2 q}{4\pi^2} \frac{d\Omega_m}{2\pi} G^{\downarrow} (\mathbf{k}_F - \mathbf{q}, -\Omega_m) G^{\downarrow} (\mathbf{k}_F + \mathbf{q} + \delta\mathbf{k}, -\Omega_m) \chi_{\uparrow \downarrow} (q, \Omega_m)
    \chi_{\downarrow \uparrow} (\mathbf{q} + \delta\mathbf{k}, -\Omega_m)                                                                                                                                                                                               \\
    = {}                                & - U^2 \int\frac{d^2 q}{4\pi^2} \frac{d\Omega_m}{2\pi} G^{\downarrow} (\mathbf{k}_F - \mathbf{q}, -\Omega_m) G^{\downarrow} (\mathbf{k}_F + \mathbf{q} + \delta\mathbf{k}, -\Omega_m) \chi_{\uparrow \downarrow} (q, \Omega_m)
    \chi_{\uparrow \downarrow } (\mathbf{q} + \delta\mathbf{k}, \Omega_m)
    \label{eq:gamma_2a_definition}
\end{align}
\end{widetext}
We emphasize that both propagators have the same sign of $i\Omega_m$ in their pole structure.
Notice the overall minus sign in the right hand side of \cref{eq:gamma_2a_definition}.

\begin{figure}
    \includegraphics[width=0.8\linewidth]{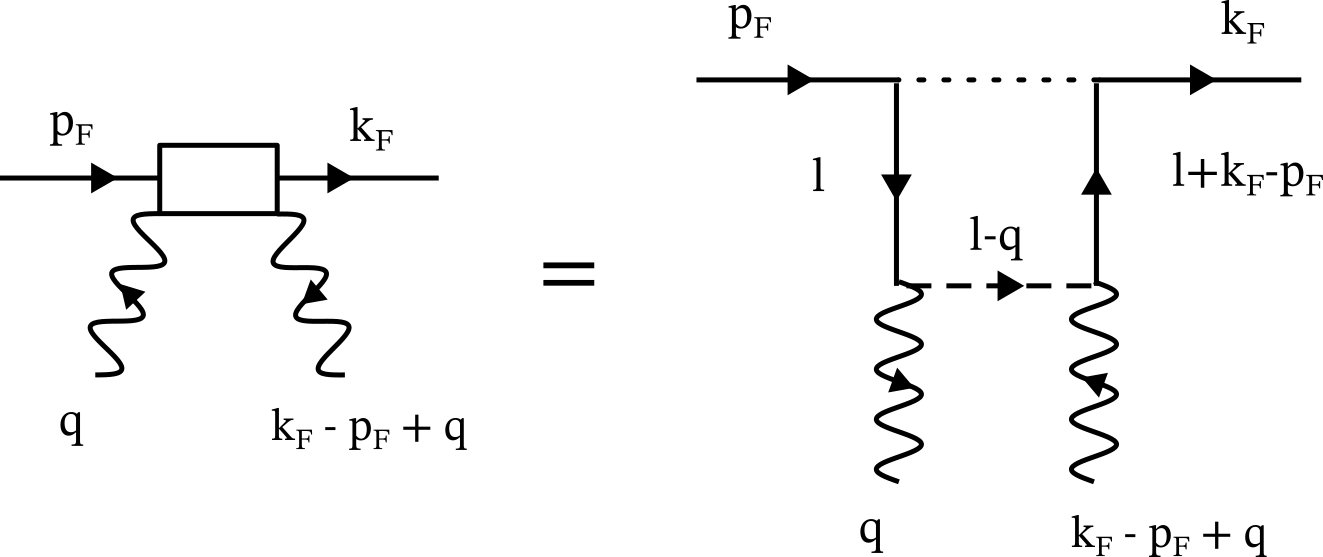}
    \caption{Vertex for the interaction between two spin-up fermions and two magnons.
        \label{fig:2magnon-vertex}}
\end{figure}

\begin{figure}
    \includegraphics[width=0.8\linewidth]{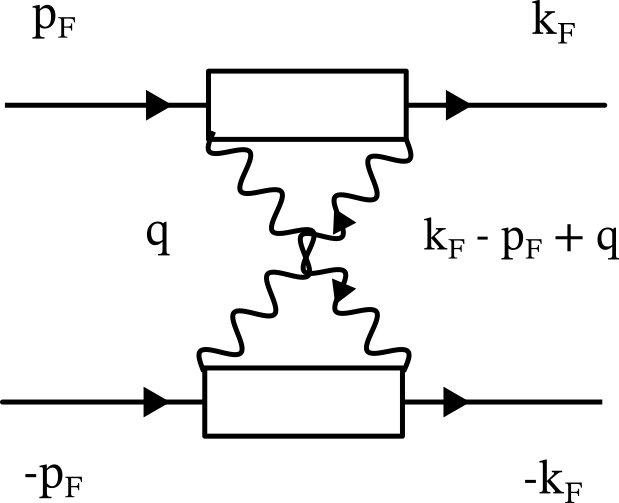}
    \caption{Contribution $\Gamma_{2b} (\delta\mathbf{k})$ to the effective interaction between two spin-up fermions from two two-magnon vertices $\Gamma_{\text{tm}}$.
        \label{fig:2magnon-2magnon}}
\end{figure}

Next, a simple experimentation shows that there also  exists a direct
coupling between two low-energy spin-up excitations
and \textit{two} magnons.
We show the corresponding vertex
$\gamma_{\text{2mag}} (\mathbf{q}, \Omega, \delta\mathbf{k})$ in \cref{fig:2magnon-vertex}.
In explicit form
\begin{multline}
    \gamma_{\text{2mag}} (\mathbf{q}, \Omega, \delta\mathbf{k})
    = -U \int\frac{d^2 l}{4\pi^2} \frac{d\omega_n}{2\pi}\\
    \times
    G^{\uparrow} (\mathbf{l},\omega_n) G^{\downarrow} (\mathbf{l} -\mathbf{q}, \omega_n-\Omega_m)
    G^{\uparrow} (\mathbf{l} + \delta\mathbf{k},\omega_n)
    \label{eq:gamma_tm_vertex}
\end{multline}
Combining the two vertices $\gamma_{\text{2mag}}$ and two magnon propagators, we obtain the effective interaction between low-energy spin-up fermions, shown in \cref{fig:2magnon-2magnon},
which we label $\Gamma_{2b} (\delta\mathbf{k})$,
\begin{align}
    \Gamma_{2b} (\delta\mathbf{k}) = {} & - U^2 \int\frac{d^2 q}{4\pi^2} \frac{d\Omega_m}{2\pi} \gamma^2_{\text{2mag}} (\mathbf{q}, \Omega, \delta\mathbf{k}) \\
                                        & \times \chi_{\uparrow \downarrow} (q, \Omega_m)
    \chi_{\uparrow \downarrow } (\mathbf{q} + \delta\mathbf{k}, \Omega_m).
    \label{eq:gamma_2b_definition}
\end{align}
\begin{figure}
    \includegraphics[width=0.8\linewidth]{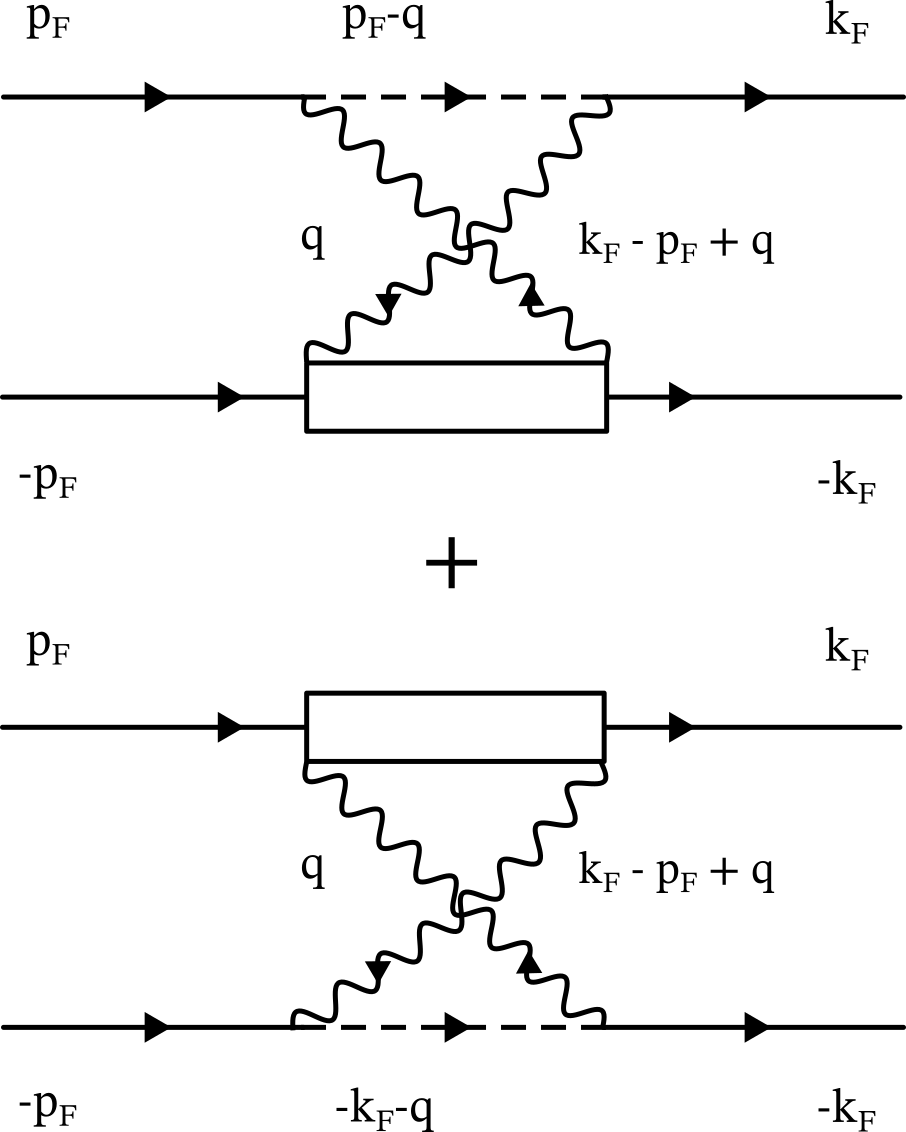}
    \caption{Contribution $\Gamma_{2c} (\delta\mathbf{k})$ to the effective interaction between two spin-up fermions from
        the process involving one
        two-magnon vertex $\gamma_{\text{2mag}}$ and two single-magnon vertices.\label{fig:2magnon-1magnon}}
\end{figure}
The overall sign in the right hand side of \cref{eq:gamma_2b_definition} is again minus.
And, finally, there exist two cross-terms involving one two-magnon vertex and two single-magnon vertices.
The corresponding diagrams are shown in \cref{fig:2magnon-1magnon}.
We label this contribution  $\Gamma_{2c} (\delta\mathbf{k})$,
\begin{multline}
    \Gamma_{2c} (\delta\mathbf{k}) =  U^2 \int\frac{d^2 q}{4\pi^2} \frac{d\Omega_m}{2\pi} \gamma_{\text{2mag}}(\mathbf{q}, \Omega, \delta\mathbf{k})\\
    \times  \left( G^{\downarrow} (\mathbf{k}_F - \mathbf{q}, -\Omega_m)+ G^{\downarrow} (\mathbf{k}_F + \mathbf{q} + \delta\mathbf{k}, -\Omega_m)\right)\\
    \times \chi_{\uparrow \downarrow} (q, \Omega_m)
    \chi_{\uparrow \downarrow } (\mathbf{q} + \delta\mathbf{k}, \Omega_m).
    \label{eq:gamma_2c_definition}
\end{multline}
Notice the overall plus sign in the right hand side of \cref{eq:gamma_2c_definition}.
Combining all three contributions, we obtain for the full 4-fermion interaction mediated by two magnon propagators
\begin{multline}
    \Gamma_{2,\text{tot}} (\bm{\delta\mathbf{k}}) =  -U^2 \int\frac{d^2 q}{4\pi^2} \frac{d\Omega_m}{2\pi}
    S\left[\mathbf{q}, \Omega_m, \bm{\delta\mathbf{k}}\right] \\
    \times \chi_{\uparrow \downarrow} (q, \Omega_m)
    \chi_{\uparrow \downarrow } (\mathbf{q} + \bm{\delta\mathbf{k}}, \Omega_m)
    \label{eq:gamma_2tot_definition}
\end{multline}
where
\begin{widetext}
\begin{equation}
    \begin{gathered}
        S\left[\mathbf{q}, \Omega_m, \bm{\delta\mathbf{k}}\right] =
        \left(\gamma_{\text{2mag}} (\mathbf{q}, \Omega, \bm{\delta\mathbf{k}})  + \frac{1}{2} \left( G^{\downarrow} (\mathbf{k}_F - \mathbf{q}, -\Omega_m)+ G^{\downarrow} (\mathbf{k}_F + \mathbf{q} + \bm{\delta\mathbf{k}}, -\Omega_m)\right)\right)^2 \\
        - \frac{1}{4} \left( G^{\downarrow} (\mathbf{k}_F - \mathbf{q}, -\Omega_m)- G^{\downarrow} (\mathbf{k}_F + \mathbf{q} + \bm{\delta\mathbf{k}}, -\Omega_m)\right)^2
    \end{gathered}
    \label{eq:S_definition}
\end{equation}
\end{widetext}

We discuss the form of $S\left[\mathbf{q}, \Omega_m, \bm{\delta\mathbf{k}}\right]$ below,
but first we analyze the structure of the equation for the pairing vertex $\phi (\mathbf{k}_F)$.
The linearized equation (the equation for the pairing scale $T^*$ defined as in the previous section) is
\begin{equation}
    \phi (\mathbf{k}) = -  \int \frac{d^2 {\mathbf{p}}}{(2\pi)^2}
    \phi (\mathbf{p}) \Gamma_{2,\text{tot}} (\mathbf{k} - \mathbf{p})
    \label{eq:gap_equation_a}
\end{equation}
Restricting to momenta near $k_F$ and performing a conventional frequency summation and integration over fermionic dispersion, we re-express (\ref{eq:gap_equation_a}) as the integral over the Fermi surface:
\begin{equation}
    \phi (\mathbf{k}_F) = - L \nu \int \frac{d {\mathbf{p}_F}}{2\pi k_F}
    \phi (\mathbf{p}_F) \Gamma_{2,\text{tot}} (\mathbf{k}_F - \mathbf{p}_F)
    \label{eq:gap_equation}
\end{equation}
where $L = \log\left(\Lambda/T^*\right)$,
$\Lambda \sim \mu_0$ is the upper cutoff of the integration transverse to the Fermi surface, and
the integration is along the Fermi surface.
As usual, fermion statistics demand that the pairing vertex be antisymmetric under fermion exchange.
In our case, the pairing involves only spin-up fermions,
hence the pairing vertex must be odd parity $\phi(\mathbf{k}_F) = -\phi(-\mathbf{k}_F)$.
In this situation, the momentum-independent part of $\Gamma_{2,\text{tot}}$ cancels out in the gap equation.  With this in mind, we
may instead work with the simplified gap equation
\begin{equation}
    \phi (\mathbf{k}_F) = - L \nu \int \frac{d \delta \mathbf{k}}{2\pi
        k_F}
    \phi(\mathbf{k}_F - \delta\mathbf{k} ) {\bar \Gamma}_{2,\text{tot}} (\delta\mathbf{k})
    \label{eq:gap_equation_simplified}
\end{equation}
where  ${\delta \mathbf{k}}$ is the variation of $\mathbf{k}_F - \mathbf{p}_F$ along the Fermi surface and $\bar{\Gamma}$ is the interaction {\it with the momentum independent piece subtracted}.

The expression for ${\bar \Gamma}_{2,\text{tot}} (\delta\mathbf{k})$, \cref{eq:gamma_2tot_definition,eq:S_definition}, is valid only at $|\delta\mathbf{k}| \leq k_F$ as
it describes the interaction mediated by long-wavelength magnons.
To proceed with this approach, we conjecture that ${\bar \Gamma}_{2,\text{tot}} (\delta\mathbf{k}) = {\bar \Gamma}_{2,\text{tot}} (0) \Psi (\delta\mathbf{k})$, where $\Psi (\delta\mathbf{k})$ is a decreasing function of
$|\delta\mathbf{k}|$ with a characteristic scale $\delta\mathbf{k}^{(0)} \leq k_F$. With this in mind, we
can estimate $T^*$ from \cref{eq:gap_equation_simplified}
by approximating $\phi (\mathbf{k}_F + \delta\mathbf{k})$ by $\phi (\mathbf{k}_F)$, $\bar{\Gamma}_{2,\text{tot}} (\delta\mathbf{k})$ by $\bar{\Gamma}_{2,\text{tot}} (0)$ and $\int \frac{d \delta \mathbf{k}}{2\pi}$ by $\delta\mathbf{k}^{(0)} /\pi$.
The equation for $T^*$ then reduces to
\begin{equation}
    1 = \lambda_{sc} L,
\end{equation}
where
\begin{equation}
    \lambda_{sc} =  -
    \nu {\bar \Gamma}_{2,\text{tot}} (0) \frac{\delta\mathbf{k}^{(0)} }{\pi k_F}.
    \label{eq:lambda_sc_definition}
\end{equation}

The scale $T^*$ is finite if $\bar{\Gamma}_{2,\text{tot}} (0) <0$, i.e., if the magnon-mediated interaction with the constant part subtracted is negative.
As is customary for an interaction peaked at small momentum transfer, the coupling $\lambda_{sc}$ has almost the same value for all odd-parity channels ($p$-wave, $f$-wave, etc) (see. e.g. [\onlinecite{Lederer2015,*Schattner2016,*Lederer2017,Klein2018,*Klein2019}]).
To differentiate between channels, one has to analyze the full
dynamical
structure of  $\bar{\Gamma}_{2,\text{tot}} (\delta\mathbf{k})$.
This will most likely select $p$-wave as the leading instability.

We now compute $\bar{\Gamma}_{2,\text{tot}} (0)$.
We have from \cref{eq:gamma_2tot_definition}
\begin{equation}
    \Gamma_{2,\text{tot}} (0) =  -U^2 \int\frac{d^2 q}{4\pi^2} \frac{d\Omega_m}{2\pi}
    S\left[\mathbf{q}, \Omega_m, 0\right]  \chi^2_{\uparrow \downarrow} (q, \Omega_m).
    \label{eq:gamma_2tot_zero}
\end{equation}
To get ${\bar \Gamma}_{2,\text{tot}} (0)$, we must remove the constant
contribution from $\Gamma_{2,\text{tot}} (0)$.
For this purpose, it is convenient to express $S$ as the sum of two terms,
\begin{equation}
    S\left[\mathbf{q}, \Omega_m, 0 \right]  =  (S_a \left[\mathbf{q}, \Omega_m\right])^2 - (S_b \left[\mathbf{q}, \Omega_m,\right])^2,
    \label{eq:S_components}
\end{equation}
where (cf. \cref{eq:S_definition}),
\begin{equation}
    \begin{aligned}
        S_a \left[\mathbf{q}, \Omega_m\right]=  & \gamma_{\text{2mag}} (\mathbf{q}, \Omega)                                                                                               \\
        +                                       & \frac{1}{2} \left( G^{\downarrow} (\mathbf{k}_F - \mathbf{q}, -\Omega_m)+ G^{\downarrow} (\mathbf{k}_F + \mathbf{q}, -\Omega_m)\right)  \\
        S_b \left[\mathbf{q}, \Omega_m\right] = & \frac{1}{2} \left( G^{\downarrow} (\mathbf{k}_F - \mathbf{q}, -\Omega_m)- G^{\downarrow} (\mathbf{k}_F + \mathbf{q}, -\Omega_m)\right).
        \label{eq:S_a_b_components}
    \end{aligned}
\end{equation}
We note that both $S_a$ and $S_b$ vanish at $q=\Omega =0$. Indeed,
$S_b \left[0, 0\right]$ vanishes identically, while for $S_a \left[0,0\right]$ the calculation of the convolution of two $G^\uparrow$ and one $G^\downarrow$ in \cref{eq:gamma_tm_vertex} yields $\gamma_{\text{2mag}} (0,0) = 1/(2\mu_0 c)$.
Combining with
$G^{\downarrow} (\mathbf{k}_F, 0) = -1/(2\mu_0 c)$, we find
$S_a \left[0,0\right]= \gamma_{\text{2mag}} (0,0) +  G^{\downarrow} (\mathbf{k}_F, 0) =0$.
This result is entirely expected~\cite{Watanabe2014}
because the same two-magnon/two-fermion interaction can be  used to analyze the renormalization of the magnon propagator $\chi_{\uparrow \downarrow } (\mathbf{q}, \Omega_m)$ by fermions.
The vanishing of $S\left[0, 0, 0\right]$ then
implies that the interaction with fermions
vanishes in the long-wavelength limit and no mass is generated for the Goldstone boson.
This is a ferromagnet-specific realization of the Adler principle (\emph{Adler zero}) for a Goldstone boson~\cite{Adler1965,Vasiliou2024}.

\begin{figure}[htb]
    \centering
    \includegraphics[width=\linewidth]{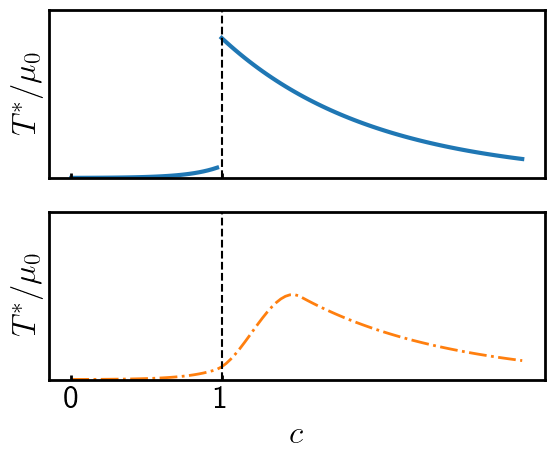}
    \caption{(Color Online) Upper panel: Schematic depiction of the pairing scale $T^*$ as a function of $c$ in both the paramagnetic ($c<1$) and ferromagnetic ($c>1$) phases for a two-dimensional system.
        Lower panel: Heuristic depiction of the pairing scale $T^*$ as a function of $c$ in a three-dimensional system,
        assuming that the magnetization quickly saturates.
        The dashed vertical line shows the location of the Stoner transition $c=1$.}
    \label{fig:Tc_vs_c}
\end{figure}

For finite frequency and momentum, we expand $S_a$ and $S_b$ in $\Omega$ and $q$, average over the angle between $\mathbf{k}_F$ and $\mathbf{q}$,and find
\begin{gather}
    S^2_a \left[\mathbf{q}, \Omega_m\right] = \frac{(i\Omega_m + \frac{q^2}{2m} \frac{c-1}{c})^2}{(2\mu_0 c)^4} = \frac{\chi^{-2}_{\uparrow \downarrow } (\mathbf{q}, \Omega_m)}{(2\mu_0 c)^2} \label{eq:S_a_expansion} \\
    S^2_b \left[\mathbf{q}, \Omega_m\right] = \frac{q^2}{2m} \frac{4\mu_0}{(2\mu_0 c)^4} \left( 1 - \frac{2i \Omega_m}{\mu_0 c} + O(q^2)\right)
    \label{eq:S_b_squared}
\end{gather}
We see that the contribution from $S^2_a$ to $\Gamma_{2,\text{tot}} (0)$ in \cref{eq:gamma_2tot_zero} reduces to a constant which has to be
subtracted from $\bar{\Gamma} (0)$ in \cref{eq:lambda_sc_definition}
\footnote{A word of caution. We checked the cancellation between $S_a \left[\mathbf{q}, \Omega_m\right]$ and $\chi_{\uparrow \downarrow } (\mathbf{q}, \Omega_m)$ at $\delta\mathbf{k}=0$. We conjecture that this holds also at a finite $\delta\mathbf{k}$.  One argument for such a cancellation is that integration over Matsubata frequency in \cref{eq:gamma_2tot_zero} runs over an infinite range $(-\infty,+\infty)$, hence  $\int d \Omega_m S^2_a \left[\mathbf{q}, \Omega_m\right]\chi^2_{\uparrow \downarrow } (\mathbf{q}, \Omega_m)$ is formally  infinite, while the true pairing interaction must indeed stay finite.}.
The contribution to $\bar{\Gamma}_{2,\text{tot}} (0)$ then comes only from
$S^2_b$. The leading term in $S^2_b$ scales as $q^2$ but is purely static.
A static  contribution to  $\Gamma_{2,\text{tot}} (0)$ in (\ref{eq:gamma_2tot_zero}) vanishes after integration over $\Omega_m$ because of the double pole in $\chi^2_{\uparrow \downarrow } (\mathbf{q}, \Omega_m)$.
A non-zero contribution
comes from the subleading term that contains $i\Omega_m$.
Substituting this term into \cref{eq:gamma_2tot_zero} and using
\begin{equation}
    \int \frac{d \Omega_m}{2\pi} \frac{1}{i\Omega_m + \frac{q^2}{2m} \frac{c-1}{c}} = \frac{1}{2}
    \label{eq:omega_integral}
\end{equation}
independent on $q$, we obtain an \emph{attractive} pairing interaction for odd-parity superconductivity in the form
\begin{equation}
    {\bar \Gamma}_{2,\text{tot}} (0) = - \left(\frac{U}{\mu_0 c^2}\right)^2 \int^{q_{max}}_0 \frac{q dq}{2\pi} \frac{q^2}{2m}
    \label{eq:gamma_2tot_attractive}
\end{equation}
The upper limit of this integration $q_{max}$ is comparable to $\delta\mathbf{k}^{(0)}$
(one has to compute $\bar{\Gamma}_{2,\text{tot}} (\delta\mathbf{k})$ to see this)
\footnote{The momentum integral can, in principle,  be explicitly evaluated if in $S^2_b$ we keep the $q$-dependence of the Green's functions next to $2\mu_0 c$. However, this calculation goes beyond the level of our approximations.}.
Using $U \nu =c$ and plugging into \cref{eq:lambda_sc_definition}, we obtain the dimensionless coupling
\begin{equation}
    \lambda_{sc} =  \frac{
        4}{\pi c} \left(\frac{\delta\mathbf{k}^{(0)}}{k_F}\right)^5.
    \label{eq:lambda_sc_final}
\end{equation}
and
\begin{equation}
    T^*  \sim \mu_0  e^{-1/\lambda_{sc}}.
    \label{eq:lambda_sc_final_1}
\end{equation}
We recall that $\delta\mathbf{k}^{(0)}/k_F \leq 1$, hence the
dimensionless coupling has no parametric smallness.  This implies that the attraction is rather strong, i.e., the pairing scale  $T^*$  in the ferromagnetically ordered state is a sizable fraction of $\mu_0 = E_F$.
While the computation of the exact value of $\lambda_{sc}$ is beyond the scope of our approach, we nevertheless emphasize that this $T^*$ is much larger in the ferromagnetic phase than in the paramagnetic phase,
particularly  near the border to ferromagnetism (cf. \cref{fig:tpnormal}).
As the systems moves deeper in to the ferromagnetic state, $c$ increases, and the
pairing scale gets smaller (i.e., superconducting $T_c \sim T^*$ falls).  We show the dependence of $T^*$ on $c$ in \cref{fig:Tc_vs_c}.
We expect superconducting $T_c$ to follow the same trend, although the transition in the paramagnetic phase will likely be first order~\cite{Chubukov2003}.

The discontinuity in $T^*$ at $c=1$ is a consequence of the first order nature of the Stoner transition in two dimensions.
For a three dimensional system, the Stoner transition is second order, and the we expect the pairing scale $T^*$ to be continuous at $c=1$.
Nonetheless, if the system rapidly saturates there should still be a pronounced peak in $T^*$ near $c=1$ as shown in the lower panel of \cref{fig:Tc_vs_c}.

\section{Conclusion}
\label{sec:conclusion}

In this work, we have considered magnon-mediated superconductivity in a two-dimensional itinerant electron system near a ferromagnetic transition.
We have shown that the nature of the pairing interaction is qualitatively different in the ferromagnetic and paramagnetic states, leading to a large discrepancy in pairing scales between the two.
In the paramagnetic state, for purely parabolic electronic dispersion, $T_c$ for pairing via paramagnon exchange is suppressed parametrically by a factor $b k^2_F$, related to the weak dispersion of
a paramagnon.
A weak paramagnon dispersion also gives rise to a topologically non-trivial  gap function with the sign change on the Matsubara axis.

The situation is qualitatively different in the ferromagnetic phase.
The fully spin-polarized nature of the ground state, which emerges discontinuously at the Stoner transition, means that
the pairing occurs between fermions with the same spin projection.
The
magnon-mediated pairing interaction involves the exchange of two transverse Goldstone magnons.
Despite the fact that the vertex for fermion-magnon scattering vanishes in the long wavelength (as expected from the Adler principle) the
pairing interaction remains attractive and sizable in the odd-parity channels, with the $p$-wave channel likely being the strongest.
We estimated the corresponding dimensionless coupling, \cref{eq:lambda_sc_final},
and fund $T_c$ as a
fraction of $E_F$ even when $b=0$ and there is no superconductivity in the paramagnetic phase.
We thus expect a  jump in the pairing temperature as the Stoner transition is crossed,
to a much larger $T_c$ in the ferromagnetic state.

The results of this study are also applicable to multi-layer graphene structures BBG, RTG and R5G, in which
it is widely  believed that superconductivity  emerges inside a ferromagnetically ordered state.  In these two-valley systems, small Fermi pockets are located near $K$ and $K'$ points in the Brillouin zone. The pairing we consider here
emerges due to intra-valley Hubbard-like interaction, and the paired fermions are located on the opposite sides of the same  Fermi surface near either $K$ or $K'$. From a general point of view, this pairing is a pair-density-wave  (PDW) phenomenon. A conventional superconductivity with zero total momentum of a pair is a pairing between one fermion near $K$ and one near $K'$.  In BBG and RTG, superconductivity has been detected in a ferromagnetic half-metal state, where there exists Fermi surfaces for  spin-up fermions in both valleys.  In this situation, both a conventional superconductivity and a PDW order are possible.  Our scenario, applied to these systems, describes a PDW order  and is a competing scenario to the ones for a conventional superconductivity in a FM state ~\cite{Dong2024,Raines2025}.
In R5G, superconductivity has been detected in a quarter-metal state which is believed to have both a FM order and valley polarization. In this situation, only spin-up fermions from one valley have a Fermi surface. Pairing of these fermions necessarily leads to a PDW order. The results of our analysis are fully applicable to this case.  Different mechanisms of PDW order in a quarter metal have been recently proposed in Refs.~[\onlinecite{Chou2025,*Geier2024,*Yang2024,*Qin2024,*Jahin2025,*Gil2025,*Dong2025,*Gaggioli2025,*Yang2024, *Parramartinez2025,*Kim2025,*Christos2025}].

\begin{acknowledgments}
We acknowledge with thanks useful discussions with Erez Berg, Piers Coleman, Zhiyu Dong, Hart Goldman, Alex Kamenev, Patrick Lee, Stevan Nadj-Perge, Pavel Nosov, Alex Thompson and  Andrea Young.
The work by AVC was supported by the National Science Foundation grant NSF: DMR-2325357.
\end{acknowledgments}

\bibliography{references_1v}

\appendix

\section{Paramagnon Propagator}\label{app:para-pi}

In the paramagnetic state, the spin polarization bubble is given by
\begin{multline}
    \Pi(q) = - T \sum_{k}  \frac{1}{i\omega_{n} + i\Omega_{m} - \xi_{\mathbf{k}+\mathbf{q}}}\frac{1}{i \omega_{n}-\xi_{\mathbf{k}}}\\
    = \sum_{\mathbf{k}}  \frac{n_{\mathbf{k}+\mathbf{q}} - n_{\mathbf{k}}}{i\Omega_{m} - \frac{q^{2}}{2m} - \frac{\mathbf{k} \cdot \mathbf{q}}{m}},
    \label{eq:piq}
\end{multline}
the two-dimensional Lindhard function on the Matsubara axis~\cite{Stern1967,Giuliani2008,Mihaila2011}.

In the zero-temperature limit, shifting $\mathbf{k}\to\mathbf{k}-\mathbf{q}$ in the first term yields
\begin{multline}
    \Pi(q)
    =
    \int^{k_{F}}_{0} \frac{kdk}{2\pi}\oint \frac{d\theta}{2\pi}\\
    \times
    \left(
    \frac{1}{i\Omega_{m} + \frac{q^{2}}{2m} - \frac{\mathbf{k} \cdot \mathbf{q}}{m}}
    - \frac{1}{i\Omega_{m} - \frac{q^{2}}{2m} - \frac{\mathbf{k} \cdot \mathbf{q}}{m}}
    \right)\\
    =
    \int^{k_F}_{0} \frac{kdk}{2\pi}\oint \frac{d\theta}{2\pi}
    \left(
    \frac
    {1}{i\Omega + \frac{q^2}{2m}- \frac{kq}{m}\cos\theta}\right.\\
    \left.
    - \frac{1}{i\Omega - \frac{q^2}{2m}- \frac{kq}{m}\cos\theta}
    \right).
\end{multline}
Factoring out $\frac{kq}{m}$ from the denominators
\begin{multline}
    \Pi(q)
    =  \frac{\nu}{q}
    \int^{k_F}_{0} dk\oint \frac{d\theta}{2\pi}\\
    \times
    \left(
    \frac{1}{i\frac{m\Omega}{kq} + \frac{q}{2k}- \cos\theta}
    - \frac{1}{i\frac{m\Omega}{kq} - \frac{q}{2k}- \cos\theta}
    \right)\\
    =  \frac{\nu}{q}
    \int^{k_F}_{0} dk \oint \frac{d\theta}{2\pi}\\
    \times
    \left(
    \frac{1}{i\frac{m\Omega}{kq} + \frac{q}{2k}- \cos\theta}
    + \frac{1}{-i\frac{m\Omega}{kq} + \frac{q}{2k}- \cos\theta}
    \right),
\end{multline}
where we have used $\theta \to \theta + \pi$ in the second term.
Defining
\begin{equation}
    a(\Omega) = \frac{q}{2} + i\frac{m\Omega}{q},
\end{equation}
we can perform the angular integration using \cref{eq:Iu} to get
\begin{equation}
    \Pi(q)
    =  \frac{\nu}{q}
    \int^{k_F}_{0} dk
    \sum_\pm
    \frac{1}{\sqrt{\frac{q}{2k} \pm i\frac{m\Omega}{q k}- 1}\sqrt{\frac{q}{2k} \pm i\frac{m\Omega}{q k}+ 1}}.
\end{equation}
Performing the momentum integration using \cref{eq:Jr} we get
\begin{multline}
    \Pi(q)
    = -\frac{\nu}{q}
    \sum_\pm
    \left.
    \sqrt{\frac{q}{2} \pm i \frac{m\Omega}{q}- k}\sqrt{\frac{q}{2} \pm i \frac{m\Omega}{q}+ k}
    \right|^{k_F}_{0}\\
    = -\frac{\nu}{q}
    \sum_\pm
    \left(
    k_F\sqrt{\frac{q}{2k_F} \pm \frac{i\Omega}{v_F q} - 1}
    \sqrt{\frac{q}{2k_F} \pm \frac{i\Omega}{v_F q} + 1}\right.\\
    \left.
    - \frac{q}{2} \mp \frac{im \Omega}{q}
    \right)\\
    = \nu - \frac{2k_F\nu}{q}\Re\left(
    \sqrt{\frac{q}{2k_F} + \frac{i\Omega}{v_F q} - 1}
    \sqrt{\frac{q}{2k_F} + \frac{i\Omega}{v_F q} + 1}
    \right).
\end{multline}
We can separate this into a static piece and a dynamic piece
\begin{equation}
    \Pi(q) = \Pi(0, \mathbf{q}) + \delta\Pi(\Omega, \mathbf{q})
\end{equation}
with
\begin{gather}
    \Pi(0, \mathbf{q}) = \nu,\\
    \begin{multlined}
        \delta\Pi(\Omega, \mathbf{q}) =
        - \frac{2k_F\nu}{q}\\
        \times\Re\left(
        \sqrt{\frac{q}{2k_F} + \frac{i\Omega}{v_F q} - 1}
        \sqrt{\frac{q}{2k_F} + \frac{i\Omega}{v_F q} + 1}
        \right).
    \end{multlined}
\end{gather}

\section{Irreducible pairing interaction}\label{app:pi}
\begin{figure}[htp]
    \centering
    \includegraphics[width=\linewidth]{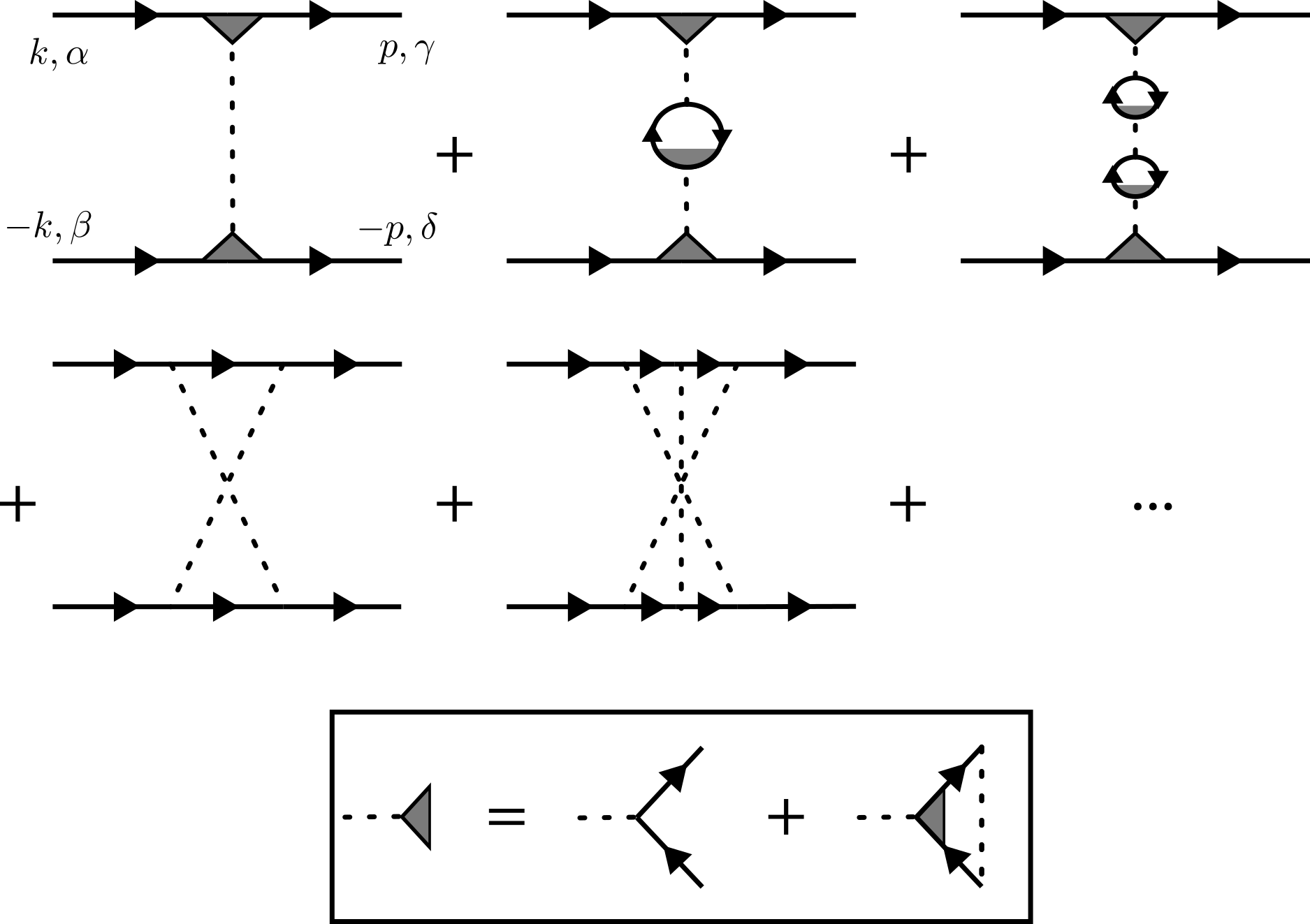}
    \caption{
        The full set of ladder and bubble diagrams contributing to the irreducible interaction in the pairing channel.
        The antisymmetrized $\Gamma_{\alpha \beta;\gamma,\delta} (k,p)$ is obtained by adding with the overall minus sign the same set of  diagrams but with $(p,\gamma)$ interchanged with $(-p,\delta)$. }
    \label{fig:spin-interaction-vertex_1}
\end{figure}

Here we present the expression for the dressed antisymmetrized interaction  $\Gamma_{\alpha \beta;\gamma,\delta} (k,p)$, which we obtain by  keeping only the diagrams with the polarization bubbles which do not depend on the internal running momenta. There are two such polarizations, $\Pi (k-p)$ and $\Pi (k+p)$. In the main text we
neglected the terms with $\Pi (k+p)$, in line with earlier works~\cite{Maiti2013,Dong2023a,Dong2023b}.  Here we keep both terms.

The diagrammatic series for the dressed interaction before anti-symmetrization are shown in Fig. \ref{fig:spin-interaction-vertex_1}. They contain the series of maximally crossed diagrams and the series of bubble diagrams with fully dressed vertices.
The series of maximally crossed diagrams sum up into
\begin{equation}
    \delta_{\alpha\gamma} \delta_{\beta \delta}
    \left( \frac{U}{1- U \Pi (k+p)} -U \right).
\end{equation}
The series of bubbles sum into
\begin{equation}
    \delta_{\alpha\gamma} \delta_{\beta \delta} \frac{U \gamma^2}{1 + 2U \gamma \Pi (k-p)}.
\end{equation}
The vertex $\gamma$ is obtained by summing up the ladder series of vertex renormalizations:
\begin{equation}
    \gamma = \frac{1}{1 - U \Pi (k-p)}.
\end{equation}
Combining maximally crossed diagrams and bubbles, we obtain after a simple algebra
\begin{equation}
    \delta_{\alpha\gamma} \delta_{\beta \delta}  U \left(\frac{1}{1 -U^2 \Pi^2 (k-p)} + \frac{1}{1-U \Pi (k+p)} -1\right)
\end{equation}
Adding the antisymmetrized piece, we obtain
\begin{equation}
    \Gamma_{\alpha \beta;\gamma,\delta} (k,p) = \Gamma^{(1)} \delta_{\alpha\gamma}\delta_{\beta \delta}  - \Gamma^{(2)} \delta_{\alpha \delta} \delta_{\beta \gamma} \end{equation}
where
\begin{align}
     & \Gamma^{(1)} = U \left(\frac{1}{1 -U^2 \Pi^2 (k-p)} + \frac{1}{1-U \Pi (k+p)} -1\right) \nonumber \\
     & \Gamma^{(2)} = U \left(\frac{1}{1 -U^2 \Pi^2 (k+p)} + \frac{1}{1-U \Pi (k-p)} -1\right).
\end{align}
Splitting this $\Gamma$ into charge and spin parts, as we did in the main text:
\begin{equation}
    \Gamma_{\alpha \beta; \gamma \delta} (k,p) = \Gamma^{ch} (k,p) \delta_{\alpha \gamma} \delta_{\beta \delta} + \Gamma^{sp} (k,p) \boldsymbol{\sigma}_{\alpha \beta} \cdot \boldsymbol{\sigma}_{\gamma \delta},
    \label{eq:gamma-dressed_1}
\end{equation}
we obtain
\begin{equation}
    \begin{aligned}
        \Gamma^{ch} (k,p)  = {} & \frac{U}{2} \left(\frac{1}{1+ U \Pi (k-p)} + \frac{1+2 U \Pi (k+p)}{1- U^2 \Pi^2 (k+p)} -1\right), \\
        \Gamma^{sp} (k,p)  = {} & - \frac{U}{2} \left(\frac{1}{1- U \Pi (k-p)} + \frac{1}{1- U^2 \Pi^2 (k+p)} -1\right).
        \label{eq:gamma-dressed-componenets_1}
    \end{aligned}
\end{equation}
One can straightforwardly check that for spin-singlet pairing (pairing vertex proportional to $i \sigma^y$), the effective interaction given by (\ref{eq:gamma-dressed_1}) and (\ref{eq:gamma-dressed-componenets_1}) is repulsive. For spin-triplet vertex (e.g., pairing vertex proportional to $\sigma^x$), the effective interaction that appears in the right hand side of the gap equation is
\begin{align}
     & -\frac{U}{2} \left(\frac{1}{1- U \Pi (k-p)} - \frac{1}{1+ U \Pi (k-p)}\right) \nonumber \\
     & + \frac{U}{2} \left(\frac{1}{1- U \Pi (k+p)} - \frac{1}{1+ U \Pi (k+p)}\right).
    \label{t_1}
\end{align}
Because the gap function in the spin-triplet channel is spatially odd, the terms with $\Pi (k-p)$ and $\Pi (k+p)$ give equal contributions. The effective pairing interaction can then be re-expressed as
\begin{equation}
    -U \left(\frac{1}{1- U \Pi (k-p)} - \frac{1}{1+ U \Pi (k-p)}\right)
    \label{t_1_1}
\end{equation}
We see that this effective interaction is negative (attractive) when $U >0$.   It is larger by the factor of 2 than the result obtained by keeping only terms proportional to $\Pi (k-p)$.

\section{Pairing interaction in the \texorpdfstring{$p-$wave}{p-wave} channel at small \texorpdfstring{$bk^2_F$}{bk2F} }
\label{app:b=0}

In this appendix we present the expressions for the pairing interaction
$\tilde{\Gamma}^{sp}_{l=1} (\Omega_m)$ in \cref{eq:para-gap-equation}
in the $p-$wave channel at the smallest $bk^2_F$.
\begin{widetext}
The generic expression for this interaction, valid for arbitrary $bk^2_F$ and arbitrary $c <1$ is
\begin{equation}
    \tilde{\Gamma}^{sp}_{l=1} (\Omega_m) =
    -\frac{c}{2\pi} \int_0^\pi \theta d \theta \frac{\cos{\theta} \sin{\theta/2}}{1-c + 4 b k^2_F \sin^3{\theta/2} - c \Im \sqrt{\cos^2\frac{\theta}{2} + \frac{\Omega^2_m}{4 v^2_F k^2_F \sin^3(\theta/2)}- i \frac{\Omega_m}{k_F v_F}}}
\end{equation}
Below we set $c=1$, i.e., consider the pairing interaction immediately before a FM instability.

For $b=0$, we have
\begin{equation}
    \tilde{\Gamma}^{sp}_{l=1} (\Omega_m) =
    \frac{1}{2\pi} \int_0^\pi  \theta d \theta \frac{\cos\theta \sin{\theta\over 2}}{\Im \sqrt{\cos^2{\theta\over 2} + \frac{\Omega^2_m}{4 v^2_F k^2_F \sin^2{\theta\over 2}}- i \frac{\Omega_m}{k_F v_F}}}
    \label{eq:p-wave-b0}
\end{equation}
\end{widetext}

\begin{figure}[htp]
    \centering
    \includegraphics[width=\linewidth]{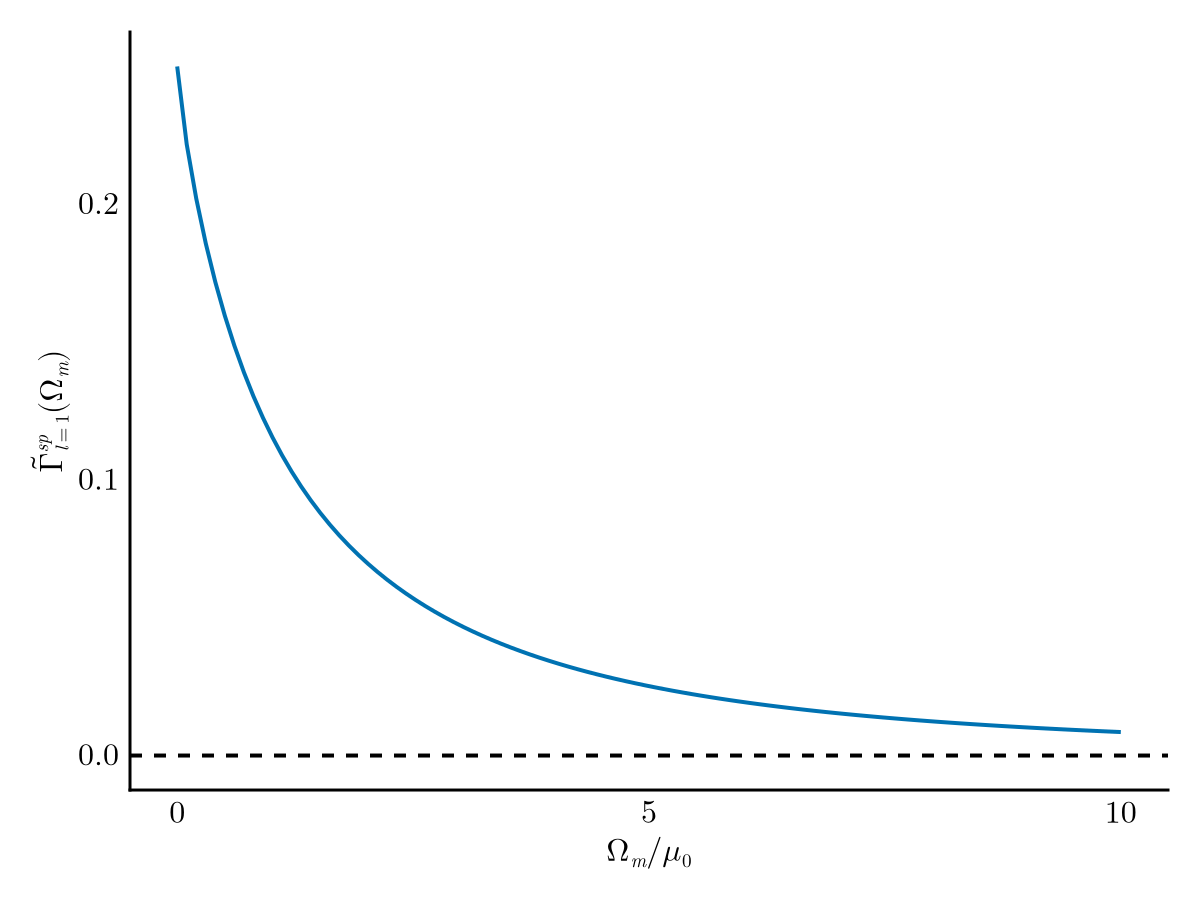}
    \caption{
        The pairing interaction $\tilde{\Gamma}^{sp}_{l=1} (\Omega_m)$ for $b=0$.
        It remains repulsive at all frequencies.
    }
    \label{fig:p-wave-b0}
\end{figure}
At the smallest $\Omega_m$, the dominant contribution comes from $\theta \approx \pi$, where $\cos(\theta/2)$ is small.
Expanding to the leading order in  $\epsilon = \pi-\theta$ and evaluating the integral over $\epsilon$, we obtain
$\tilde{\Gamma}^{sp}_{l=1} (0)  =1/4$.
At large $\Omega_m$,
\begin{multline}
    \Im \sqrt{\cos^2{\theta\over2} + \frac{\Omega^2_m}{4 v^2_F k^2_F \sin^2{\theta\over2}}- i \frac{\Omega_m}{k_F v_F}} \\
    = \sin{\theta\over2} + O((k_F v_F/\Omega_m))^2.
\end{multline}
Substituting into
\cref{eq:p-wave-b0} and integrating over $\theta$, we obtain $\tilde{\Gamma}^{sp}_{l=1} (\Omega_m) = 0.5 (2v_F k_F/\Omega_m)^2$.
We see that in the limits of small and large frequencies, the $p-$wave component of the pairing interaction is positive, i.e., repulsive.
We show the full $\tilde{\Gamma}^{sp}_{l=1} (\Omega_m)$ at $b=0$ in \cref{fig:p-wave-b0}.
We see that it is repulsive at all frequencies.
\begin{figure}[htp]
    \centering
    \includegraphics[width=\linewidth]{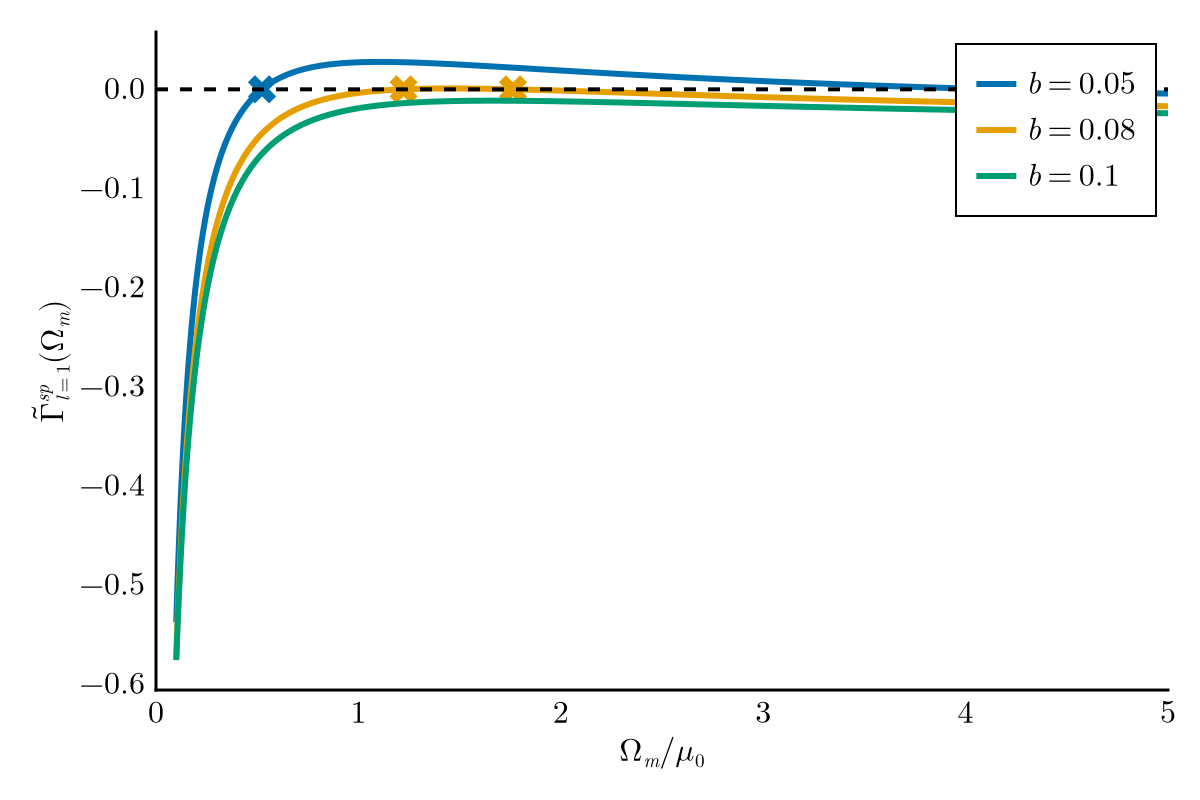}
    \caption{
        The pairing interaction $\tilde{\Gamma}^{sp}_{l=1} (\Omega_m)$ for different values of $bk_F^2$.
        The interaction is attractive at small and large frequencies and repulsive at intermediate frequencies.
        Cross marks indicate the points at which $\tilde{\Gamma}^{sp}_{l=1} (\Omega_m)$ changes sign.
    }
    \label{fig:p-wave-b-nonzero}
\end{figure}

\begin{figure}[htp]
    \centering
    \includegraphics[width=\linewidth]{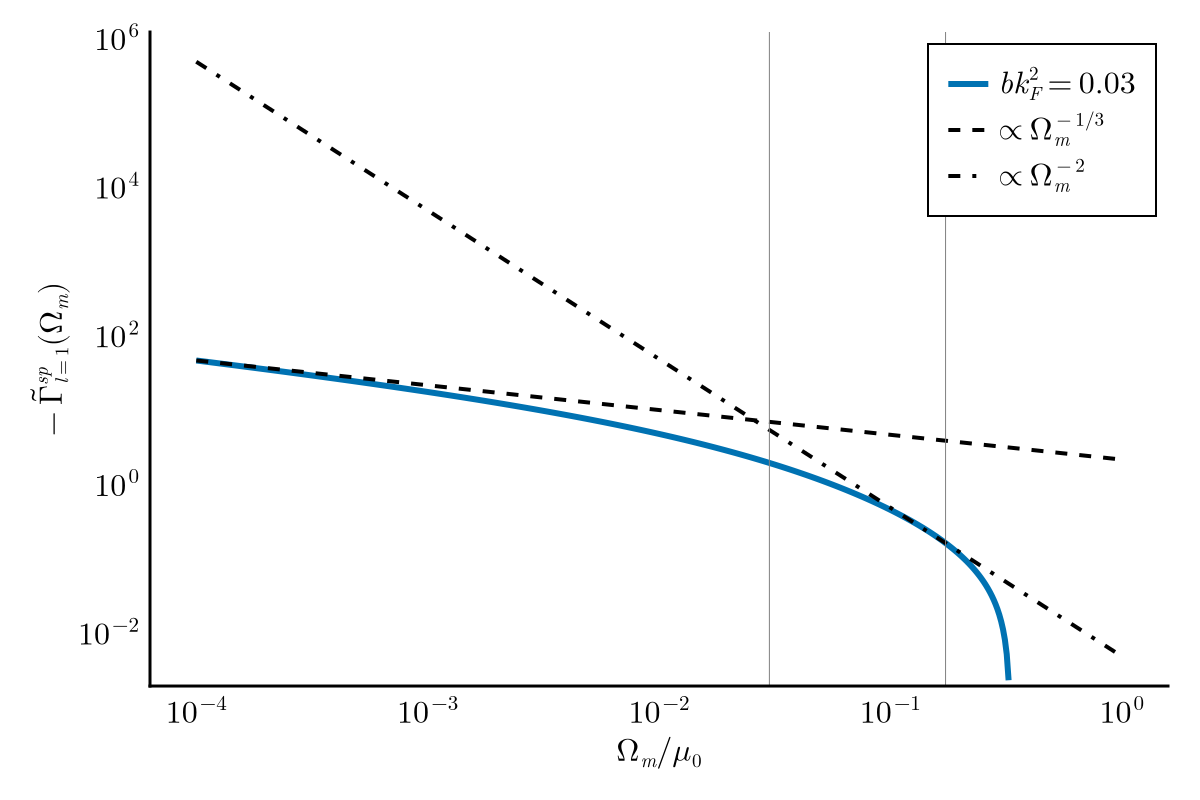}
    \caption{
        Log-log plot of the pairing interaction $\tilde{\Gamma}^{sp}_{l=1} (\Omega_m)$ along with the power law scaling for the smallest and intermediate frequencies.
        Thin ertical lines indicate the crossover scales $(bk^2_F)^{1/2}$ and $bk^2_F$.
    }
    \label{fig:p-wave-powers}
\end{figure}

The situation changes at a finite $b$.
Now at large $\Omega_m$,  $\tilde{\Gamma}^{sp}_{l=1} (\Omega_m) = - (bk^2_F)/2$ is negative, i.e., attractive.
The same happens at the smallest $\Omega_m$.
The two leading contributions here come from $\theta$ near zero and near $\pi$.
Combining these contributions, we find that $\tilde{\Gamma}^{sp}_{l=1} (\Omega_m)$ is again negative  and scales as $(bk^2_F)/\Omega^2_m$  at $\Omega_m$ in between $bk^2_F$ and $(bk^2_F)^{1/2}$ and as $1/((bk^2_F)^2|\Omega_m|)^{1/3}$  at $|\Omega_m| <(k^2_F)$.
We show the full $\tilde{\Gamma}^{sp}_{l=1} (\Omega_m)$ in \cref{fig:p-wave-b-nonzero}.
We see that at small $bk^2_F$ it is attractive at small and large frequencies and repulsive at intermediate frequencies.
As $bk^2_F$ increases, the range where
$\tilde{\Gamma}^{sp}_{l=1} (\Omega_m)$ is positive (repulsive), shrinks and eventually vanishes.

\begin{figure}[htp]
    \centering
    \includegraphics[width=\linewidth]{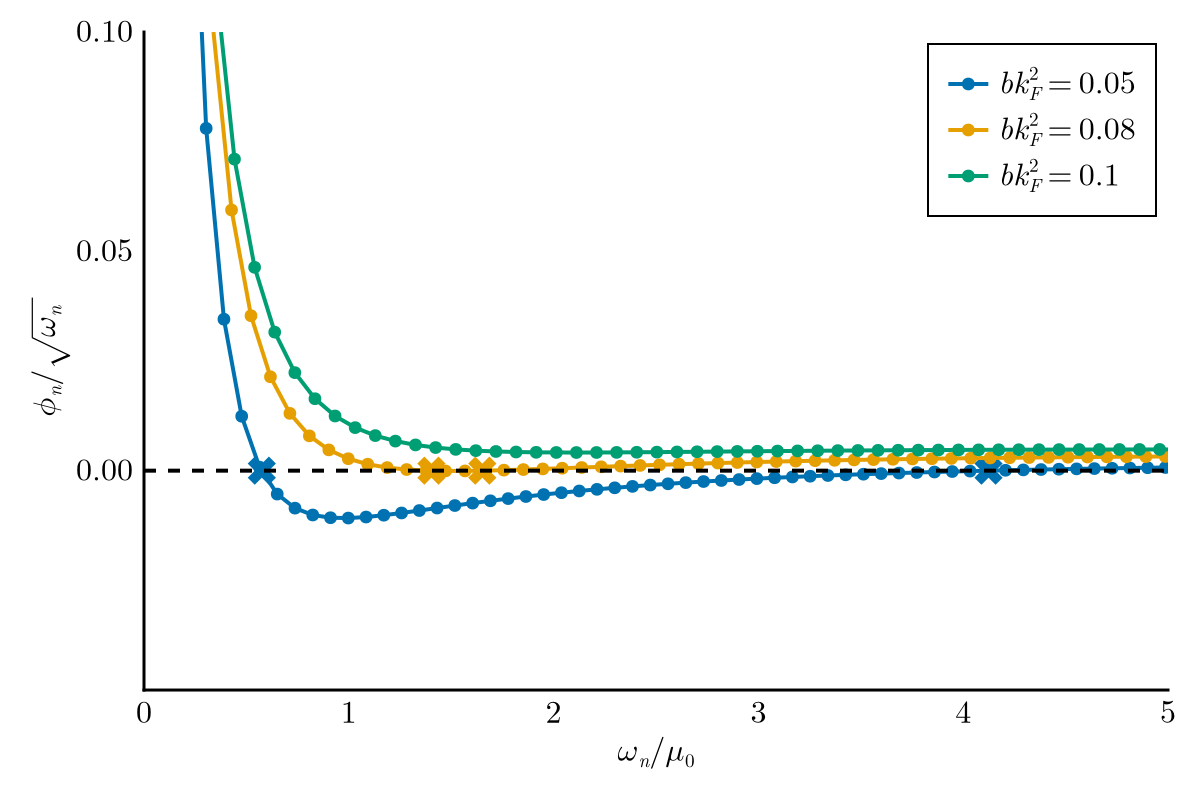}
    \caption{
        The gap function $\phi_n$ for different values of $bk_F^2$.
        Cross-marks indicate the points at which $\phi_n$ changes sign.
        As $b$ increases, grow closer together and at $b_c$ collide and move into the complex plane.
    }
    \label{fig:gap-function}
\end{figure}

In \cref{fig:gap-function} we show the gap function $\phi_n$ - the solution of the gap equation \cref{eq:para-gap-equation} at
the onset temperature $T^*$, at different $b$.
The temperature $T^*$ is non-zero for any non-zero $b$  and evolves with $b$ as in \cref{fig:Tc_vs_c} in the main text.
We  see that at small $bk^2_F$, $\phi_n$ changes sign two times as a function of Matsubara frequency.
As we said in the main text, the points at which $\phi_n$ changes sign is the center of a dynamical vortex.
As $bk^2_F$ increases, the two vortices come close to each other and merge
at some finite $b$.
At larger $b$ they move in opposite directions away from the Matsubara axis and eventually leave the upper half-plane of frequency.

\section{Useful Integrals}\label{sec:useful-integrals}

In the course of our calculations, we make use of the following integrals
\begin{gather}
    I[u] \equiv \oint \frac{d\theta}{2\pi} \frac{1}{u - \cos\theta}
    = \frac{1}{\sqrt{u+1}\sqrt{u-1}}
    \label{eq:Iu}\\
    J[r] \equiv \int \frac{kdk}{\sqrt{r-k}\sqrt{r+k}}
    = -\sqrt{r+k}\sqrt{r-k} + C
    \label{eq:Jr}
\end{gather}
which are derived in the following subsections.

\subsection{\texorpdfstring{$I[u]$}{I[u]}}

We can perform the integral \cref{eq:Iu} by contour integration with the change of variables $z = e^{i\theta}$.
Then,
\begin{multline}
    I[u] \equiv \oint \frac{d\theta}{2\pi} \frac{1}{u - \cos\theta}\\
    = \oint_\gamma \frac{dz}{2\pi i z} \frac{1}{u -\frac{z+z^{-1}}{2}}\\
    = -2\oint_\gamma \frac{dz}{2\pi i} \frac{1}{z^2-2u z+1}
    ,
\end{multline}
where the path $\gamma$ is the unit circle which has poles at $z = u \pm \sqrt{u^2-1}$.
Clearly the integral has branch cut when $\Im u =0$ and $\Re u \in [-1, 1]$.
We can thus perform the integration for a particular region of $u$ and analytically continue around to branch cut.
Taking $\sqrt{\cdots}$ to have a branch cut along the negative real axis, as is conventional, we therefore find that
\begin{equation}
    I[u] = \frac{1}{\sqrt{u+1}\sqrt{u-1}}.
\end{equation}

\subsection{\texorpdfstring{$J[r]$}{J[r]}}

For \cref{eq:Jr} we rewrite
\begin{multline}
    J[r] = \int \frac{kdk}{\sqrt{r-k}\sqrt{r+k}}
    = \int dk \frac{r+k - (r-k)}{2\sqrt{r-k}\sqrt{r+k}}\\
    = \int dk \left(
    \frac{\sqrt{r+k}}{\sqrt{r-k}}
    - \frac{\sqrt{r-k}}{\sqrt{r+k}}
    \right)
    = -\sqrt{r+k}\sqrt{r-k} + C
\end{multline}

\end{document}